\documentclass[aps,pre,twocolumn,groupedaddress]{revtex4-1}
\usepackage{graphicx}
\usepackage{dcolumn}
\usepackage{bm}
\usepackage{amsmath}
\usepackage{amssymb}
\usepackage{color}

\begin{document}

\title{Vesicle budding induced by binding of curvature-inducing proteins}

\author{Hiroshi Noguchi}
\email[]{noguchi@issp.u-tokyo.ac.jp}
\affiliation{Institute for Solid State Physics, University of Tokyo, Kashiwa, Chiba 277-8581, Japan}

\date{\today}

\begin{abstract}
Vesicle budding induced by protein binding that generates an isotropic spontaneous curvature
is studied using a mean-field theory.
Many spherical buds are formed via protein binding.
As the binding chemical potential increases,
the proteins first bind to the buds and then to the remainder of the vesicle.
For a high spontaneous curvature and/or high bending rigidity of the bound membrane,
it is found that a first-order transition occurs between a small number of large buds and a large number of small buds.
These two states coexist around the transition point.
The proposed scheme is simple and easily applicable to  many interaction types, so we investivate
 the effects of inter-protein interactions, the protein-insertion-induced changes in area,
the variation of the saddle-splay-modulus, and the area-difference-elasticity energy.
The differences in the preferred curvatures for curvature sensing and generation are also clarified.
\end{abstract}

\maketitle

\section{Introduction}

In living cells, biomembranes have various shapes depending on their functions.
Spherical vesicles, which are involved in endo/exocytosis and vesicle transports,
are formed through membrane budding.
Many types of protein are known to be involved in such budding~\cite{mcma05,suet14,joha15,bran13,hurl10,mcma11}.
For example, clathrins assemble on a membrane and form a spherical bud with  $\sim 100$-nm diameter;
the clathrin-coated bud is then pinched off by dynamin and other proteins~\cite{hurl10,mcma11,schm11,kaks18,avin15}.

To understand the bud formation mechanism,
many experiments have been conducted {\it in vitro} using giant liposomes.
Buds can be induced by the area difference elasticity (ADE) in single-phase liposomes~\cite{seif97,svet09,hota99,saka12,holl21},
and by the separation of liquid-ordered and liquid-disordered phases in three-component liposomes~\cite{baum03,baci05,yana08}.
Moreover, budding induced by
 polymer anchoring~\cite{tsaf01} and binding of proteins such as clathrin~\cite{sale15} and annexins~\cite{boye18}
has been observed.
A contrasting mechanism involves Bin/Amphiphysin/Rvs (BAR) superfamily proteins, which bend the membrane 
in one direction along its axis
 than in the other direction; thus, their binding generates cylindrical membrane tubes rather than spherical buds~\cite{mcma05,suet14,joha15,itoh06,mim12a}.
These curvature-inducing proteins also sense the local membrane curvature and
exhibit preferred binding to membranes with their preferred curvatures.
To investigate this curvature sensing, 
a liposome with a narrow membrane tube pulled by optical tweezers and a micropipette~\cite{baum11,sorr12,prev15,rosh17}
and different sizes of liposomes~\cite{zeno19} have been employed for various types of proteins.

Vesicle shape transformation has been numerically investigated  using various types of coarse-grained membrane models.
Budding has been simulated using dynamically triangulated membranes~\cite{kuma01,kohy03,nogu15a,peze19,tame20}, 
meshless membranes~\cite{nogu17a},  
dissipative particle dynamics~\cite{baga09,naka18}, phase-field models~\cite{lowe09}, and so on.
Spontaneous tubule formation has been simulated
by taking the membrane anisotropic spontaneous curvature into account~\cite{nogu16,nogu17a,nogu19a,rama18}.
However, in contrast with experiments,
 the simulated vesicle size is limited by the computational costs.
In simulations, the vesicle-to-bud size ratio is typically less than $10$, which is far less
 than that in the experiments on budding induced by protein binding, i.e., $\sim 100$.

In previous theoretical analyses of the budding~\cite{lipo92,sens03,fore14,frey20},
a single bud in a flat membrane was typically considered;
that is, the remaining vesicle area was regarded as a membrane reservoir.
However,
to understand the global vesicle shape in thermal equilibrium, the entire vesicle must be considered.

This study examines changes in vesicle shape in response to increased binding chemical potential using a mean-field theory.
The entire vesicle is treated explicitly using a simplified geometry.
Hence, the binding 
dependence on the membrane properties and various interactions between membrane and proteins and between proteins
are clarified.

Protein binding to an arbitrarily shaped vesicle is described in Sec.~\ref{sec:sens}.
The budding of a vesicle with a constant membrane area is examined in Sec.~\ref{sec:bud1},
with the effects of the saddle-splay-modulus difference and  inter-protein interactions being presented 
in Secs.~\ref{sec:results2} and \ref{sec:results3}, respectively.
The effects of the area changes due to the protein insertion 
and the ADE energy are examined  in Secs.~\ref{sec:area} and \ref{sec:ade},
respectively.
Finally, the discussion and summary are presented in Secs.~\ref{sec:dis} and \ref{sec:sum},
respectively.

\section{Local protein binding}\label{sec:sens}

Binding of proteins or other molecules to a membrane is considered, because this
 binding modifies the membrane bending rigidity and spontaneous curvature. 
First, the case where the membrane area is unchanged by the binding is considered;
that is, the proteins adsorb on the membrane surface, 
or the hydrophobic segments of the protein inserted into the membrane are so
small that the area change is negligible.
The free energy $F$ of a vesicle consists of the bending energy $F_{\rm cv}$, 
binding energy,
 inter-protein interaction energy, and mixing entropy:
\begin{eqnarray}\label{eq:F0}
F &=&  F_{\rm cv} + \int dA \ \Big\{ - \frac{\mu}{a_{\rm p}}\phi   + b \phi^2 + \\ \nonumber
&&   \frac{k_{\rm B}T}{a_{\rm p}}[\phi \ln(\phi) +  (1-\phi) \ln(1-\phi) ] \Big\}, \\ \label{eq:Fcv0}
F_{\rm cv} &=&  4\pi\bar{\kappa}_{\rm d}(1-g) + \int dA \ \Big\{ 2\kappa_{\rm d}H^2(1-\phi)  \\ \nonumber
&&  +  2\kappa_{\rm p}(H-H_0)^2\phi  
+ (\bar{\kappa}_{\rm p}-\bar{\kappa}_{\rm d})C_1C_2 \phi  \Big\},
\end{eqnarray}
where $A$ is the membrane area,  $\phi$ is the local protein density ($\phi=1$ at the maximum coverage),
$\mu$ is the chemical potential of the protein binding,
 $k_{\rm B}T$ is the thermal energy, $C_1$ and $C_2$ are the principal curvatures,
and $H$ is the mean curvature of each position ($H=(C_1+C_2)/2$).
Here, subscript p indicates the bound-membrane quantities
and  $a_{\rm p}$ is the area covered by one protein
(the maximum number of the bound proteins is $A/a_{\rm p}$).
The remaining terms are defined in the below discussion.

The membrane is in a fluid phase and
 $F_{\rm cv}$ is the second-order expansion to the curvature~\cite{canh70,helf73}.
The bare (protein-unbound) membrane has bending rigidity  $\kappa_{\rm d}$ with zero  spontaneous curvature. 
The bound proteins are considered to be laterally isotropic (i.e., they have no preferred bending direction); hence,
the bound membrane has a bending rigidity $\kappa_{\rm p}$ and finite spontaneous mean curvature $H_0$,
which is the half of the spontaneous curvature $C_0$ because $H_0=C_0/2$.
Here, $H_0$ is used instead of $C_0$, because $1/H_0$ is the radius of the spherical membrane 
to minimize the bending energy of the protein-bound membrane (the second term in the integral of Eq.~(\ref{eq:Fcv0}) for $F_{\rm cv}$).
The first term of Eq.~(\ref{eq:Fcv0}) represents
the integral over the Gaussian curvature $C_1C_2$
with the saddle-splay modulus $\bar{\kappa}_{\rm d}$  (also called the Gaussian modulus)~\cite{safr94}
of the bare membrane, where $g$ is the genus of the vesicle.
This type of bending energy for protein binding has been used in Refs.~\cite{tame20,gout21},
as well as for a cylindrical membrane with protein rods of anisotropic spontaneous curvature~\cite{nogu15b}.

In this study, $\kappa_{\rm d}=20k_{\rm B}T$ and 
$\bar{\kappa}_{\rm d}/\kappa_{\rm d}=-1$ are used as they
 are typical values of lipid membranes.
Since  $\bar{\kappa}_{\rm p}$ for the bound membrane is unknown,
two cases are considered: $\bar{\kappa}_{\rm p} = - \kappa_{\rm p}$ and  $\bar{\kappa}_{\rm p} = \bar{\kappa}_{\rm d}$.
In the former case, the proteins have the same ratio to the bending rigidity as for the bare membrane.
In the latter, the proteins do not change the saddle-splay modulus of the membrane.
The former condition is considered in this work,  unless  otherwise specified
(the latter condition is examined in Sec.~\ref{sec:results2}).

\begin{figure}
\includegraphics{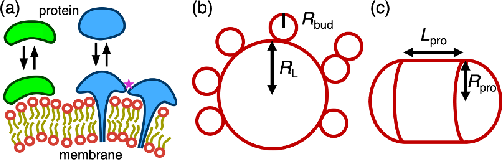}
\caption{
Schematic of the membrane.
(a) Binding and unbinding of proteins to the membrane.
(b) Budded vesicle. The buds and the large remaining vesicle have spherical shapes with radii of $R_{\rm bud}$
and  $R_{\rm L}$, respectively.
(c) Prolate vesicle approximated by cylinder combined with two hemispheres with radius $R_{\rm pro}$ and cylinder length $L_{\rm pro}$.
}
\label{fig:cart}
\end{figure}
\begin{figure}[t]
\includegraphics[]{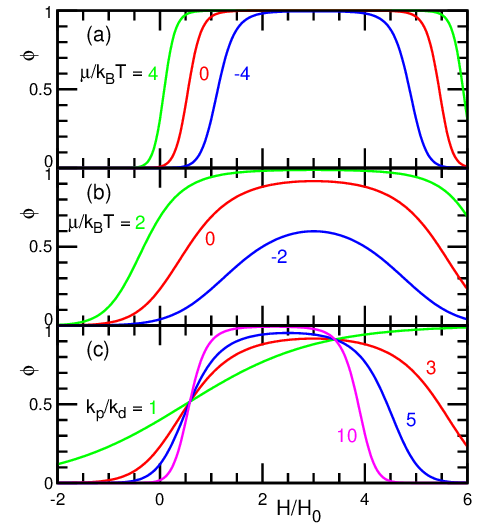}
\caption{
Protein density $\phi$ as a function of the local mean curvature $H$ for a spherical membrane ($H=C_1=C_2$).
(a), (b) The binding chemical potential $\mu$ is varied for  $\kappa_{\rm p}/\kappa_{\rm d}=3$ 
at (a) $H_0^2 a_{\rm p} = 0.04$ and (b) $H_0^2 a_{\rm p} = 0.01$.
(c) The bending rigidity $\kappa_{\rm p}$ of the bound membrane is varied for  $\mu=0$ and $H_0^2 a_{\rm p} = 0.01$.
}
\label{fig:phic1}
\end{figure}

The first term in the integral of Eq.~(\ref{eq:F0}) represents the protein binding energy.
Under higher $\mu$, more proteins bind to the membrane. 
The last two terms of Eq.~(\ref{eq:F0}) represent the pairwise inter-protein interactions and the mixing entropy of the bound proteins, respectively.
Proteins have repulsive or attractive interactions at $b>0$ and $b<0$, respectively.
In the addition of the direct inter-protein interactions, the proteins interact with each other via membrane (bending deformation, height mismatch, and Casimir-like interactions)~\cite{reyn07,auth09,aran96,four99,gout21}.
Here, we consider only weak curvature-independent interactions so that proteins are uniformly distributed in each component.
Note that membrane-mediated interactions between soft objects such as proteins can be much weaker than hard objects~\cite{nogu17}.

The inter-protein interaction via the spontaneous curvature is also considered in this work.
If the proteins have large hydrophobic domains above the membrane, as depicted on the right side of Fig.~\ref{fig:cart}(a),
a repulsive interaction yields a positive spontaneous curvature~\cite{baum11}.
Opposite (negative) curvature can be induced by an attractive interaction.
In this study, linear dependence on the protein density is considered as a leading-order approximation;
i.e., $H_0$ is replaced by $H_0'= H_0+ H_1\phi$ in Eq.~(\ref{eq:Fcv0}).
The effect of this condition on a constant-area vesicle is discussed in Sec.~\ref{sec:results3}.
For the other conditions, only constant spontaneous curvature ($H_1=0$) is considered, for simplicity.

In thermal equilibrium,
the protein density $\phi$ is locally determined for the given curvatures.
When the inter-protein interactions are negligible ($b=0$ and $H_1=0$),
 $\phi$ is obtained from $\partial F/\partial \phi|_{H}=0$ for a spherical membrane ($H=C_1=C_2$)
as a sigmoid function:
\begin{equation}\label{eq:phi}
\phi = \frac{1}{1+\exp\big[\frac{a_{\rm p}}{k_{\rm B}T}(2\kappa_{\rm dif}H^2 -4\kappa_{\rm p}H_0H + \sigma_{\rm p} )\big]},
\end{equation}
where $\kappa_{\rm dif}=\kappa_{\rm p}-\kappa_{\rm d}+(\bar{\kappa}_{\rm p}-\bar{\kappa}_{\rm d})/2$ and 
 $\sigma_{\rm p}= - \mu/a_{\rm p} + 2\kappa_{\rm p}H_0^2$.
For $\kappa_{\rm dif}=0$ and $H=0$, this expression corresponds to Eq.~(5) in Ref.~\cite{kris19}
and Eq.~(4) in Ref.~\cite{gout21}, respectively.
For a membrane with arbitrary curvature,
the first energy term in the parentheses in Eq.~(\ref{eq:phi})
is replaced by $2(\kappa_{\rm p}-\kappa_{\rm d})H^2 + (\bar{\kappa}_{\rm p}-\bar{\kappa}_{\rm d})C_1C_2$.
When inter-protein interactions exist,
$\phi$ is obtained by iteratively solving  Eq.~(\ref{eq:phi}), while adding 
$2b\phi - 8\kappa_{\rm p}(H-H_0)H_1\phi + 6 \kappa_{\rm p}H_1^2\phi^2$ within the parentheses.
For a flat membrane ($H=0$) 
with $\phi=0.5$ (i.e., the proteins bind to the half of the membrane area),
\begin{equation}\label{eq:phihalf}
 \mu= \Big[ \big(2H_0^2 + 4H_0H_1 + \frac{3}{2}H_1^2\big)\kappa_{\rm p} +b \Big]a_{\rm p}.
\end{equation}

Figure~\ref{fig:phic1} shows that the protein binding depends on the local membrane curvature.
For a high curvature of $H_0$ or high rigidity of $\kappa_{\rm p}$,
the density $\phi$ changes steeply from $0$ to $1$ with a small increase in $H$ (see Figs.~\ref{fig:phic1}(a) and (c)).
Here, $\phi$ exhibits a maximum  at $H= (\kappa_{\rm p}/\kappa_{\rm dif})H_0$
(given by $d\phi/dH=0$).
This curvature dependence is called curvature sensing.
In contrast, the free-energy minimum given by $dF/dH =0$ is 
$H= [\kappa_{\rm p}\phi/(\kappa_{\rm dif}\phi + \kappa_{\rm d})]H_0$, 
because the membrane must bend together.
Therefore, the curvature generated by the protein binding is lower than the preferred curvature for the curvature sensing,
even for $\phi=1$.
For a cylindrical membrane, the preferred curvatures for the curvature sensing and generation
are  $H= [\kappa_{\rm p}/(\kappa_{\rm p}-\kappa_{\rm d})]H_0$ and $H=\{\kappa_{\rm p}\phi/[(\kappa_{\rm p}-\kappa_{\rm d})\phi + \kappa_{\rm d}]\}H_0$, respectively.

To close this section, two variants of the bending energy for protein binding are discussed.
When proteins adhere to the membrane surface, as depicted on the left of Fig.~\ref{fig:cart}(a),
and the membrane composition beneath the proteins is unchanged by the binding,
the bending energy can be expressed as
\begin{eqnarray}\label{eq:Fcv1}
F_{\rm cv} &=&  4\pi\bar{\kappa}_{\rm d}(1-g) + \int dA \ \Big\{ 2\kappa_{\rm d}H^2 \\ \nonumber
&&   + [ 2\kappa_{\rm pa}(H-H_{\rm 0a})^2 + \bar{\kappa}_{\rm pa}C_1C_2 ]\phi  \Big\}.
\end{eqnarray}
This bending energy is identical to Eq.~(\ref{eq:Fcv0}) with
  $\kappa_{\rm p}=\kappa_{\rm pa}+\kappa_{\rm d}$, $\bar{\kappa}_{\rm p}=\bar{\kappa}_{\rm pa}+\bar{\kappa}_{\rm d}$,
 $H_0= [\kappa_{\rm pa}/(\kappa_{\rm pa}+\kappa_{\rm d})] H_{\rm 0a}$, and  $\sigma_{\rm p}= - \mu/a_{\rm p} + 2\kappa_{\rm pa}H_{\rm 0a}^2$.
$\kappa_{\rm pa}$ is the bending rigidity of protein itself,
while $\kappa_{\rm p}$ is the rigidity including the membrane beneath the protein.
For a spherical membrane,  the preferred curvatures for the curvature sensing and generation
are  $H= [\kappa_{\rm pa}/(\kappa_{\rm pa}+\bar{\kappa}_{\rm pa}/2)]H_{\rm 0a}$ and $H=\{\kappa_{\rm pa}\phi/[(\kappa_{\rm pa}+\bar{\kappa}_{\rm pa}/2 )\phi + \kappa_{\rm d}]\}H_{\rm 0a}$, respectively.
This type of $F_{\rm cv}$ with $\bar{\kappa}_{\rm pa}=0$ was used in Refs.~\cite{prev15,rosh17,frey20}.

In some previous studies~\cite{sorr12,shi15,gov18,tozz19,kris19},
the binding-induced modification of the bending rigidity was not accounted for, and
the following bending energy  was used:
\begin{equation}\label{eq:Fcv2}
F_{\rm cv} =   \int dA \ \Big\{ 2\kappa_{\rm d}(H-\phi H_{\rm 0})^2  \Big\}.
\end{equation}
This corresponds to the condition of $\kappa_{\rm p}=\kappa_{\rm d}$,  $\bar{\kappa}_{\rm p}=\bar{\kappa}_{\rm d}$, $b=2\kappa_{\rm d}H_0^2$, and $\sigma_{\rm p}= - \mu/a_{\rm p}$
in the present model.
The preferred curvatures for the curvature sensing and generation
are  $H= \infty$ and $H=\phi H_{\rm 0}$, respectively.
In their model, neighboring proteins interact via the bending energy through $b=2\kappa_{\rm d}H_0^2$. 
Thus, this term has often been neglected~\cite{rama00,shlo09}.
Since the linear and quadratic terms of $\phi$ represent membrane-protein and inter-protein interactions, respectively,
they should be separately treated.
As discussed above, the present model is generic and involves these two bending-energy models as the specific parameter sets.

\section{Budded vesicle with constant area}\label{sec:bud1}

\subsection{Free energy}\label{sec:buden}

We investigate budding of a vesicle induced by protein binding in thermal equilibrium.
The vesicle has a spherical topology ($g=0$) with no pores.
For this theoretical analysis,
the vesicle is assumed to form $n_{\rm bud}$ buds, each with radius $R_{\rm bud}$. 
The remainder of the vesicle is assumed to form a spherical shape with a radius of $R_{\rm L}$, as depicted in Fig.~\ref{fig:cart}(b), 
with $R_{\rm L}> R_{\rm bud}$.
The proteins have a positive spontaneous curvature ($H_0>0$) such that they bind to the buds more than the remaining vesicle.
In the absence of the proteins, the vesicle forms a prolate shape, which is approximated as shown in Fig.~\ref{fig:cart}(c).

In this section, we consider that the membrane maintains a constant surface area, i.e., the protein binding does not change the membrane area 
(corresponding to the left protein in Fig.~\ref{fig:cart}(a)).
The protein densities at the large spherical component of the vesicle and at the buds are $\phi_{\rm L}$ and $\phi_{\rm bud}$, respectively.
The proteins are assumed to be homogeneously distributed in each region.
The total membrane area $A= 4\pi R_0^2$ is kept constant;
thus, the total bud area is given by
\begin{equation} \label{eq:R0}
\frac{A_{\rm bud}}{4\pi} = n_{\rm bud}{R_{\rm bud}}^2 = R_0^2(1 - r^2),
\end{equation}
where $r=R_{\rm L}/R_0$. 
The volume $V$ of the vesicle is fixed:
\begin{equation} \label{eq:V0}
V = \frac{4\pi}{3} \big(R_{\rm L}^3 + n_{\rm bud}{R_{\rm bud}}^3 \big).
\end{equation}
Using Eqs.~(\ref{eq:R0}) and (\ref{eq:V0}),
the bud curvature can be expressed as 
\begin{equation} \label{eq:VR}
\frac{R_0}{R_{\rm bud}} =  \frac{1- r^2 }{V^*  -   r^3},
\end{equation}
where $V^*=V/(4\pi R_0^3/3)$ is the reduced volume.
The free energy $F$ of the vesicle is given by
\begin{eqnarray} \label{eq:F1}
F &=&  F_{\rm L} + F_{\rm bud}, \\
\frac{F_{\rm L}}{4\pi} &=&  \label{eq:FL1}
\bar{\kappa}_{\rm d} + 2(\kappa_{\rm dif}\phi_{\rm L}+\kappa_{\rm d}) -4\kappa_{\rm p}H_0R_0r\phi_{\rm L} \\ \nonumber
&& + R_0^2r^2(\sigma_{\rm p}\phi_{\rm L} + b \phi_{\rm L}^2) \\ \nonumber
&&+  \frac{ k_{\rm B}TR_0^2 r^2}{a_{\rm p}}[\phi_{\rm L} \ln(\phi_{\rm L}) +  (1-\phi_{\rm L}) \ln(1-\phi_{\rm L}) ],    \\
\frac{F_{\rm bud}}{4\pi} &=& n_{\rm bud} \label{eq:FB1}
\Big\{  2[ (\kappa_{\rm p}-\kappa_{\rm d})\phi_{\rm bud}  \\ \nonumber
&&  + (\bar{\kappa}_{\rm p}-\bar{\kappa}_{\rm d})(\phi_{\rm bud}- s_{\rm neck}\phi_{\rm L}) +\kappa_{\rm d}] \\ \nonumber
&& - 4 \kappa_{\rm p}H_0R_{\rm bud}\phi_{\rm bud} 
 + {R_{\rm bud}}^2(\sigma_{\rm p}\phi_{\rm bud} + b \phi_{\rm bud}^2 ) \\ \nonumber
&&  + \frac{k_{\rm B}T {R_{\rm bud}}^2}{a_{\rm p}}[\phi_{\rm bud} \ln(\phi_{\rm bud})  \\ \nonumber
&& +  (1-\phi_{\rm bud}) \ln(1-\phi_{\rm bud}) ]   \Big\}   \\
 &=&  (1 - r^2) \label{eq:FB2}
\Big\{  2[ (\kappa_{\rm p}-\kappa_{\rm d})\phi_{\rm bud}  \\ \nonumber
&&  + (\bar{\kappa}_{\rm p}-\bar{\kappa}_{\rm d})(\phi_{\rm bud}- s_{\rm neck}\phi_{\rm L}) 
 +\kappa_{\rm d}]\left[\frac{1-r^2}{V^* - r^3}\right]^2 \\ \nonumber
&& - 4 \kappa_{\rm p}H_0R_0 \phi_{\rm bud}\frac{1-r^2}{V^* - r^3 }  +  R_0^2(\sigma_{\rm p}\phi_{\rm bud} + b \phi_{\rm bud}^2) \\ \nonumber
&&  + \frac{k_{\rm B}TR_0^2}{a_{\rm p}}[\phi_{\rm bud} \ln(\phi_{\rm bud})  \\ \nonumber
&& +  (1-\phi_{\rm bud}) \ln(1-\phi_{\rm bud}) ]   \Big\},
\end{eqnarray}
where $F_{\rm L}$ and $F_{\rm bud}$ are the free energies for the large spherical region and buds, respectively.
In Eq.~(\ref{eq:FB2}),  $n_{\rm bud}$ is treated as a real number.
This is a reasonable assumption for $n_{\rm bud}\gg 1$.
For $n_{\rm bud}<10$, we treated $n_{\rm bud}$ as an integer,
so that $r$ is geometrically determined for each $n_{\rm bud}$,
and $F_{\rm bud}$ is calculated using Eq.~(\ref{eq:FB1}).

The neck region connecting a bud and the main spherical component (or connecting buds) has a saddle shape, with $H\simeq 0$.
Since these necks have low curvature and small area, the influence of the integral of the mean curvature is negligible.
However, it is necessary to examine the Gaussian curvature.
The second term of Eq.~(\ref{eq:FB1}) represents the integral of the Gaussian curvature of the bud and neck regions,
where $s_{\rm neck}\phi_{\rm L}$ is  the protein density of the neck region.
At $s_{\rm neck}=1$, the necks have the same protein density as the main spherical component.
This is reasonable because both regions have  $H\simeq 0$.
However, the necks have high principal curvatures ($\sim \pm 0.1$\,nm$^{-1}$),
which likely prevent large proteins from binding to them, that is, $s_{\rm neck}=0$.
In general,  $s_{\rm neck}\in [0,1]$.
Here, $s_{\rm neck}=0$ is used, unless  otherwise specified.
The neck exerts a very small influence, as discussed in Sec.~\ref{sec:results2}.

The free energy $F$ is a function of three variables $r$, $\phi_{\rm L}$, and $\phi_{\rm bud}$.
In the case of no inter-protein interactions ($b=0$ and $H_1=0$),
 $F$ is expressed as a function of one variable, i.e., $r$, using Eq.~(\ref{eq:phi}).
Hence, the free-energy minimum is numerically determined by $dF/dr=\partial F/\partial r = 0$ 
(Note that $\partial F/\partial \phi = 0$ for $\phi$ satisfying Eq.~(\ref{eq:phi})).
When the inter-protein interactions exist, Eq.~(\ref{eq:phi}) is iteratively solved.
In this study, the iterations were repeated until the difference in $\phi$ was less than $10^{-8}$.
Typically, fewer than $10$ iterations were performed
at $b\ne 0$ and $H_1R_0 \lesssim 100$.

The surface tension $\sigma_{\rm d}$ and the osmotic pressure $\Pi$ can be imposed as a Lagrange multiplier to maintain the area and volume, respectively: $\breve{F}=F_{\rm L} + F_{\rm bud} + \sigma_{\rm d}A - \Pi V$.
Then, $\sigma_{\rm d}$ and $\Pi$ are determined from $\partial\breve{F}/\partial R_{\rm L}|_{R_{\rm bud},n_{\rm bud}}=0$ and $\partial\breve{F}/\partial R_{\rm bud}|_{R_{\rm L},n_{\rm bud}}=0$:  

\begin{eqnarray} \label{eq:pi1}
\Pi &=& \frac{1}{R_{\rm L}-R_{\rm bud}}\Big\{
  4\kappa_{\rm p}H_0\frac{R_{\rm L}\phi_{\rm bud} -R_{\rm bud}\phi_{\rm L}}{R_{\rm L}R_{\rm bud}} \\ \nonumber
&& + 2\sigma_{\rm p}(\phi_{\rm L} - \phi_{\rm bud} )  + 2b({\phi_{\rm L}}^2 - {\phi_{\rm bud}}^2 )  \\ \nonumber
&& + \frac{2k_{\rm B}T}{a_{\rm p}}\big[  \phi_{\rm L} \ln(\phi_{\rm L}) -  (1-\phi_{\rm L}) \ln(1-\phi_{\rm L})  \\ \nonumber
&& - \phi_{\rm bud} \ln(\phi_{\rm bud}) +  (1-\phi_{\rm bud}) \ln(1-\phi_{\rm bud}) \big] \Big\},
\end{eqnarray}

\begin{eqnarray} \label{eq:sd1}
 \sigma_{\rm d} 
 &=& \frac{\Pi R_{\rm L}}{2} 
+ \frac{2 \kappa_{\rm p}H_0\phi_{\rm L}}{R_{\rm L}}  -  \sigma_{\rm p}\phi_{\rm L}  - b{\phi_{\rm L}}^2 \\ \nonumber
&&  - \frac{k_{\rm B}T}{a_{\rm p}}[\phi_{\rm L} \ln(\phi_{\rm L}) +  (1-\phi_{\rm L}) \ln(1-\phi_{\rm L}) ] \\ \label{eq:sd2}
 &=& \frac{\Pi R_{\rm bud}}{2} 
+ \frac{2 \kappa_{\rm p}H_0\phi_{\rm bud}}{R_{\rm bud}}  -  \sigma_{\rm p}\phi_{\rm bud}  - b{\phi_{\rm bud}}^2 \\ \nonumber
&&  - \frac{k_{\rm B}T}{a_{\rm p}}[\phi_{\rm bud} \ln(\phi_{\rm bud}) +  (1-\phi_{\rm bud}) \ln(1-\phi_{\rm bud}) ].
\end{eqnarray}
The first terms of Eqs.~(\ref{eq:sd1}) and (\ref{eq:sd2}) represent the Laplace tension.

\begin{figure}[]
\includegraphics[]{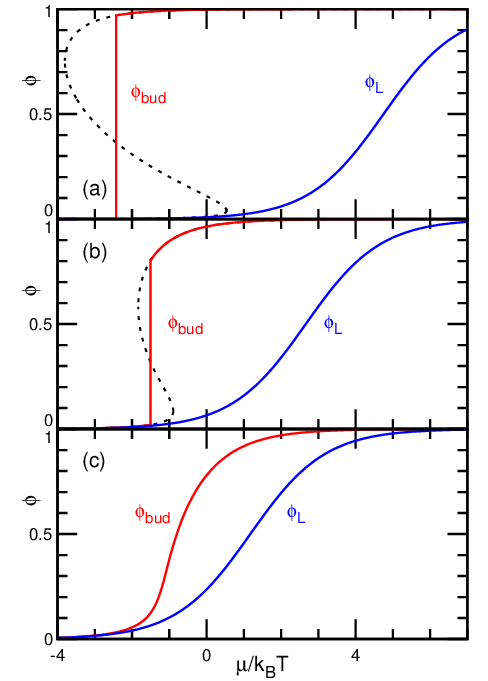}
\caption{
Chemical potential  $\mu$ dependence of the protein densities in the vesicle and buds, $\phi_{\rm L}$ and $\phi_{\rm bud}$, respectively,
for (a) $H_0R_0=200$, (b) $H_0R_0=150$, and (c) $H_0R_0=100$ at $\kappa_{\rm p}/\kappa_{\rm d}=3$, $V^*=0.9$, and $b=H_1=0$.
The solid lines represent thermal equilibrium states.
The dashed lines represent the metastable and free-energy-barrier states.
}
\label{fig:v9phi}
\end{figure}

\begin{figure}[]
\includegraphics[]{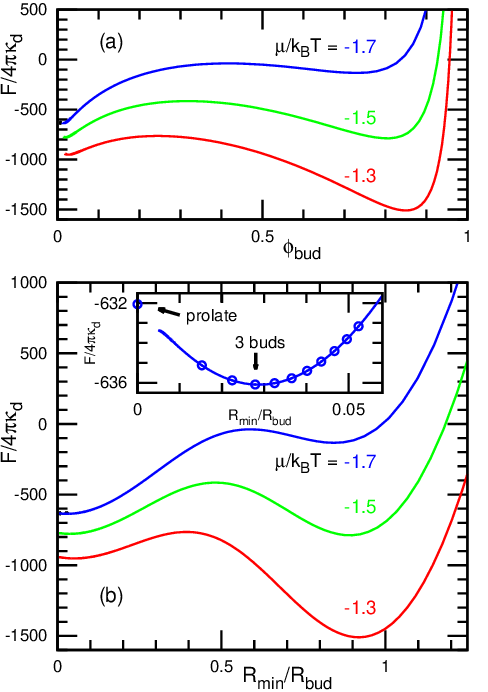}
\caption{
Free energy profiles for (a) the protein density in the buds $\phi_{\rm bud}$ and 
(b) the bud curvature $R_{\rm min}/R_{\rm bud}$
at $H_0R_0=150$, $\kappa_{\rm p}/\kappa_{\rm d}=3$, $V^*=0.9$, and $b=H_1=0$,
where $R_{\rm min}= (\kappa_{\rm dif}+\kappa_{\rm d})/\kappa_{\rm p}H_0$
 is the radius of the minimum bud curvature energy at $\phi_{\rm bud}=1$.
From top to bottom,  $\mu/k_{\rm B}T = -1.7$, $-1.5$, and $-1.3$.
The inset in (b) shows an enlarged plot at small $R_{\rm min}/R_{\rm bud}$,
in which circles represent $F$ at integer $n_{\rm bud}$.
The free energy minimum is obtained at $R_{\rm min}/R_{\rm bud}=0.0279$ ($n_{\rm bud}=3$), 
$R_{\rm min}/R_{\rm bud}=0.891$ ($n_{\rm bud}=2737$), and $R_{\rm min}/R_{\rm bud}=0.919$ ($n_{\rm bud}=2908$)
for $\mu/k_{\rm B}T = -1.7$, $-1.5$, and $-1.3$, respectively.
}
\label{fig:en}
\end{figure}

\begin{figure}[]
\includegraphics[]{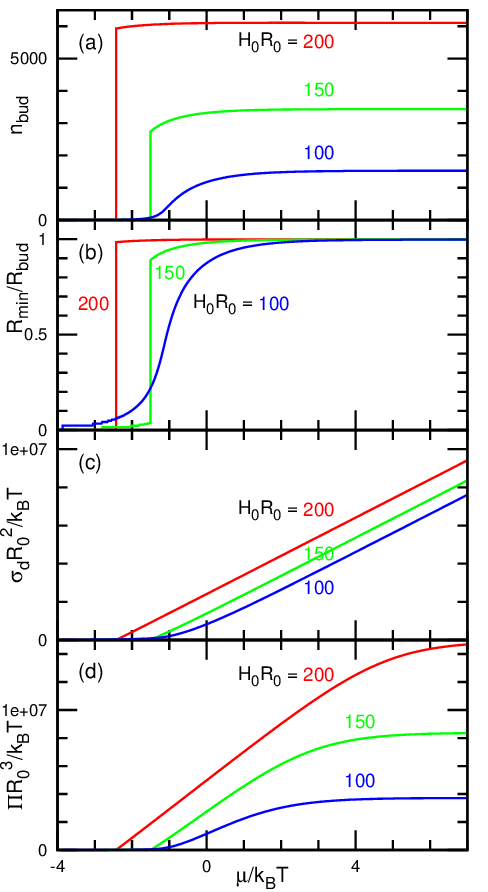}
\caption{
Chemical potential  $\mu$ dependence of (a) the bud number $n_{\rm bud}$, (b) the bud curvature $R_{\rm min}/R_{\rm bud}$, (c) the surface tension $\sigma_{\rm d}$, and (d) the osmotic pressure $\Pi$
for $H_0R_0=100$, $150$, and $200$ at $\kappa_{\rm p}/\kappa_{\rm d}=3$, $V^*=0.9$, and $b=H_1=0$.
The curvature radius of the minimum bud curvature energy is $R_{\rm min}= (\kappa_{\rm dif}+\kappa_{\rm d})/\kappa_{\rm p}H_0$ at $\phi_{\rm bud}=1$.
}
\label{fig:v9bud}
\end{figure}

A prolate vesicle is modeled with a simple geometry using a combination of one cylinder and two hemispheres, as shown in Fig.~\ref{fig:cart}(c).
The surface area and volume of the vesicle are
$A=4\pi R_{\rm pro}^2 + 2\pi R_{\rm pro}L_{\rm pro}$ and $V=(4\pi/3)R_{\rm pro}^3 + \pi R_{\rm pro}^2L_{\rm pro}$,
respectively;
thus, the reduced volume is $V^*= (1+3L_{\rm pro}/4R_{\rm pro})/(1+L_{\rm pro}/2R_{\rm pro})^{3/2}$.
The free energy is given by
\begin{eqnarray}
\frac{F_{\rm pro}}{4\pi} &=&  \label{eq:FP0}
\bar{\kappa}_{\rm d} + 2(\kappa_{\rm dif}\phi_{\rm sp}+\kappa_{\rm d}) -4\kappa_{\rm p}H_0R_{\rm pro}\phi_{\rm sp}  \nonumber \\
&& + R_{\rm pro}^2(\sigma_{\rm p}\phi_{\rm sp} + b \phi_{\rm sp}^2) \\ \nonumber
&&+  \frac{ k_{\rm B}TR_{\rm pro}^2}{a_{\rm p}}[\phi_{\rm sp} \ln(\phi_{\rm sp}) +  (1-\phi_{\rm sp}) \ln(1-\phi_{\rm sp}) ]   \\ \nonumber
&&+ \frac{L_{\rm pro}}{2}\Big\{  \frac{1}{2R_{\rm pro}}[(\kappa_{\rm p}-\kappa_{\rm d})\phi_{\rm cy}+\kappa_{\rm d})]
 - 2\kappa_{\rm p}H_0\phi_{\rm cy}  \\ \nonumber
&& + R_{\rm pro}(\sigma_{\rm p}\phi_{\rm cy} + b \phi_{\rm cy}^2 ) \\ \nonumber
&&  + \frac{k_{\rm B}T {R_{\rm pro}}}{a_{\rm p}}[\phi_{\rm cy} \ln(\phi_{\rm cy}) 
 +  (1-\phi_{\rm cy}) \ln(1-\phi_{\rm cy}) ]   \Big\},
\end{eqnarray}
where $\phi_{\rm cy}$ and $\phi_{\rm sp}$ are the protein densities in the cylindrical and spherical regions, respectively.
The expression  well approximates the energy of the prolate vesicle.
We determine the thermal equilibrium state (i.e., the lowest free energy $F$) by comparing $F$ of budded vesicles and prolate.

In this study, we consider the vesicle at $V^*> 1/\sqrt{2}\approx 0.707$,
in which a vesicle with one bud can exist ($R_{\rm bud}=R_{\rm L}$ at $V^*= 1/\sqrt{2}$).
In this range, the prolate vesicle is the most stable state in the absence of the protein.
We use  $a_{\rm p} = 100$\,nm$^2$ and $R_0= 10\,\mu$m for the area of proteins and the radius of giant liposomes, unless  otherwise specified,
i.e., $a_{\rm p}/R_0^2= 10^{-6}$.

\begin{figure}[]
\includegraphics[]{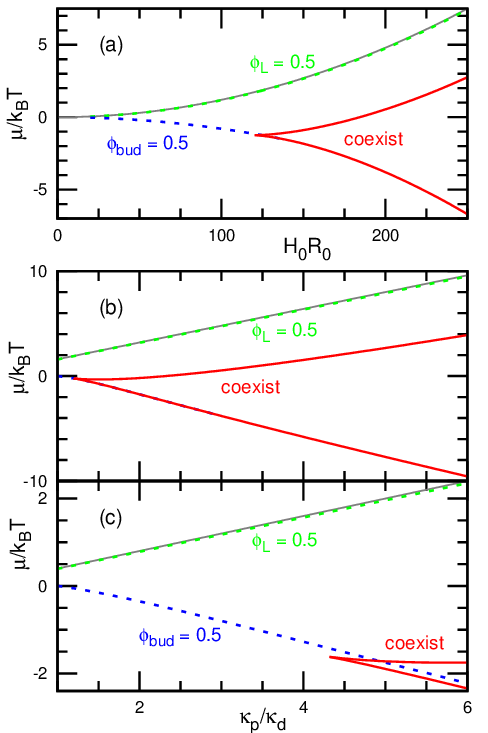}
\caption{
Phase diagrams  at $V^*=0.9$ and $b=H_1=0$.
(a)  $H_0$ vs. $\mu$ for $\kappa_{\rm p}/\kappa_{\rm d}=3$. 
(b),(c)  $\kappa_{\rm p}/\kappa_{\rm d}$ vs. $\mu$ for (b) $H_0R_0=200$  and (c) $H_0R_0=100$.
Two states with small and large bud numbers coexist in
the region between the two solid lines.
The upper and lower dashed lines represent the states with $\phi_{\rm L}=0.5$ and $\phi_{\rm bud}=0.5$, respectively.
The solid gray lines are given by Eq.~(\ref{eq:phihalf}) and well overlay the data for $\phi_{\rm L}=0.5$.
}
\label{fig:pd0}
\end{figure}

\begin{figure}[]
\includegraphics[]{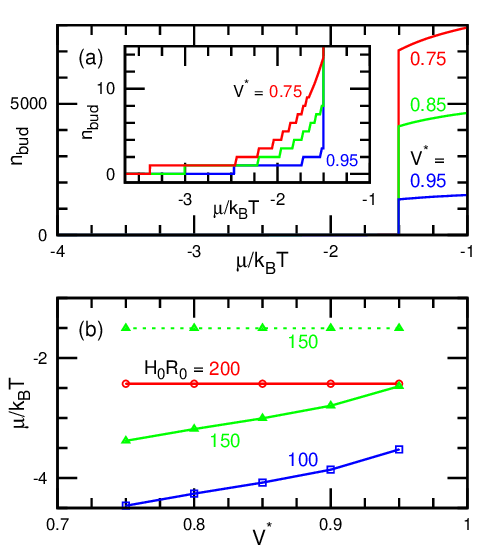}
\caption{
Dependence on reduced volume $V^*$ at $\kappa_{\rm p}/\kappa_{\rm d}=3$ and $b=H_1=0$.
(a) Number $n_{\rm bud}$ of the buds for $V^*=0.75$, $0.85$, and $0.95$ at $H_0R_0=150$.
Inset: Magnified graph for small $n_{\rm bud}$.
(b) Chemical potential  $\mu$ for the transition points for $H_0R_0=100$, $150$, and $200$.
The solid lines represent the transition from the prolate to budded vesicle,
having a single bud  for $H_0R_0=100$ and $150$, and many buds for $H_0R_0=200$.
The dashed lines represent the shape transition from low  $\phi_{\rm bud}$ with small $n_{\rm bud}$ 
to high $\phi_{\rm bud}$ with large $n_{\rm bud}$  at $H_0R_0=150$.
}
\label{fig:v}
\end{figure}

\begin{figure}[]
\includegraphics[]{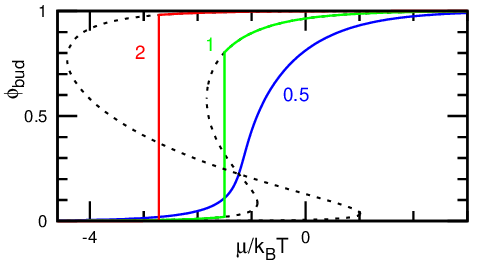}
\caption{
Dependence of $\phi_{\rm bud}$ on the area ratio $a_{\rm p}/R_0^2$
at $H_0R_0=150$, $\kappa_{\rm p}/\kappa_{\rm d}=3$, $V^*=0.9$, and $b=H_1=0$.
From top to bottom,  $a_{\rm p}/R_0^2=2\times 10^{-6}$, $10^{-6}$, and $5\times 10^{-7}$.
The dashed lines represent metastable and free-energy-barrier states.
}
\label{fig:phi_dp}
\end{figure}

\begin{figure}[]
\includegraphics[]{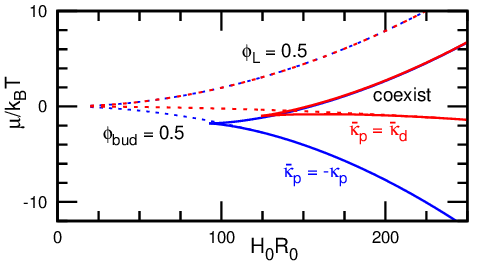}
\caption{
Effects of $\bar{\kappa}_{\rm p}$ on
the phase diagram for  at $\kappa_{\rm p}/\kappa_{\rm d}=5$, $V^*=0.9$, and $b=H_1=0$.
Data for $\bar{\kappa}_{\rm p}=\bar{\kappa}_{\rm d}$ and $\bar{\kappa}_{\rm p}=-\kappa_{\rm p}$ are shown.
Two states with a small and large number of buds coexist in
the region between the two solid lines.
The middle and lower dashed lines represent the states with $\phi_{\rm bud}=0.5$ 
at $\bar{\kappa}_{\rm p}=\bar{\kappa}_{\rm d}$ and $\bar{\kappa}_{\rm p}=-\kappa_{\rm p}$, respectively.
The upper dashed lines for $\phi_{\rm L}=0.5$ and the upper boundary of the coexistence region overlay for two cases ($\bar{\kappa}_{\rm p}=\bar{\kappa}_{\rm d}$ and $\bar{\kappa}_{\rm p}=-\kappa_{\rm p}$).
}
\label{fig:pdg}
\end{figure}

\begin{figure}[]
\includegraphics[]{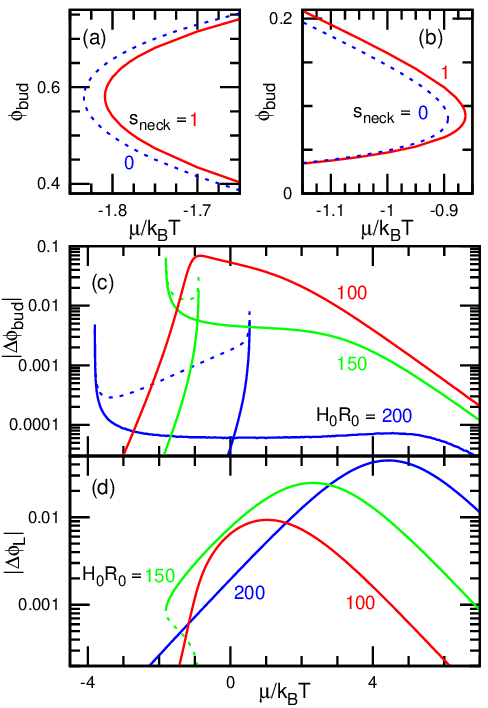}
\caption{
Effects of the relative coverage ratio $s_{\rm neck}$ of the neck region at  $\kappa_{\rm p}/\kappa_{\rm d}=3$, $V^*=0.9$, and $b=H_1=0$.
(a), (b) Enlarged graphs of the boundaries of the coexistence region for $s_{\rm neck}=0$ and $1$ at  $H_0R_0=150$.
The dashed lines ($s_{\rm neck}=0$) correspond to the data shown in Fig.~\ref{fig:v9phi}(b).
(c), (d) Differences 
in (c) $\phi_{\rm bud}$ and (d) $\phi_{\rm bud}$ for $s_{\rm neck}=0$ and $1$  
at  $H_0R_0=100$, $150$, and $200$.
The dashed lines in (c) and (d) represent free-energy-barrier states.
Here, Eq.~(\ref{eq:FB2}) is used even for $n_{\rm bud}<10$.
}
\label{fig:phi_sa}
\end{figure}

\subsection{Results}\label{sec:result0}
\subsubsection{No inter-protein interactions}\label{sec:results1}

First, we describe protein binding in the absence of the inter-protein interactions ($b=0$ and $H_1=0$).
As  $\mu$ increases,
 $\phi_{\rm bud}$ increases, and subsequently, $\phi_{\rm L}$ 
 increases, as shown in Fig.~\ref{fig:v9phi}.
At low $H_0$, $\phi_{\rm bud}$ continuously changes from $0$ to $1$ (see Fig.~\ref{fig:v9phi}(c)).
In contrast, at high $H_0$, $\phi_{\rm bud}$ exhibits a first-order transition. 
As the double minima of the free energy are shown in Fig.~\ref{fig:en}(a),
 metastable states exist around the transition point
(see also van der Waals loop depicted by the dashed lines in Figs.~\ref{fig:v9phi}(a) and (b)).
On the other hand,  $\phi_{\rm L}$ always exhibits a continuous change.
Accompanied by the discrete change in  $\phi_{\rm bud}$,
 $n_{\rm bud}$ and  $R_{\rm bud}$  also discretely change, as shown in Figs.~\ref{fig:en}(b) and \ref{fig:v9bud}.
Hence, around the transition point, 
the vesicles of small $\phi_{\rm bud}$ for buds with large $R_{\rm bud}$
coexist with those having large $\phi_{\rm bud}$ with small $R_{\rm bud}$.
At large $\mu$, $R_{\rm bud}$ reaches the radius  of curvature 
for the curvature generation $R_{\rm min}$ at $\phi_{\rm bud}=1$.
Therefore, with increasing $H_0$, the radius decreases as $R_{\rm bud} \sim 1/H_0$, 
and the bud number increases as $n_{\rm bud} \sim H_0^2$.
The free energy barrier between a small number of buds and many buds can be very high: 
$\Delta F \simeq 10^5k_{\rm d}$ and $5\times 10^5k_{\rm d}$ for $H_0R_0=150$ and $200$, respectively, at $\kappa_{\rm p}/\kappa_{\rm d}=3$ and $V^*=0.9$
(see Fig.~\ref{fig:en} for $H_0R_0=150$).
Although a free energy barrier can exist for the single-bud formation depending on conditions~\cite{lipo92,sens03,fore14,frey20,hass17},
this is much greater than the single-bud barrier, since more than $1000$ buds are formed.
Thus, the metastable states presented in this study can remain even on  much longer time scales than the single-bud formation period.

After the buds are almost covered by the proteins, 
the surface tension $\sigma_{\rm d}$ linearly increases. Further, osmotic pressure $\Pi$ increases first linearly and then saturates
(see Figs.~\ref{fig:v9bud}(c) and (d)). 
The vesicle ruptures when $\sigma_{\rm d}$ overcomes the lysis tension.
This threshold is typically $1$--$25$\, mN/m,
 depending on the membrane composition and conditions~\cite{evan00,evan03,ly04}.
Since $\sigma_{\rm d}= 10^7 k_{\rm B}T/R_0^2$ corresponds to $0.4$\,mN/m,
the lipid membranes are not yet ruptured in the range shown in Fig.~\ref{fig:v9bud}(c).
However, protein binding (for transmembrane proteins in particular) may reduce the lysis tension
and induce rupturing at a lower tension.

The dependences on  $H_0$ and $\kappa_{\rm p}/\kappa_{\rm d}$ are clearly captured by the phase diagrams shown in Fig.~\ref{fig:pd0}.
With increasing $H_0$, the middle points $\phi_{\rm bud}=0.5$ and  $\phi_{\rm L}=0.5$ shift to 
smaller and larger values, respectively. 
At $H_0R_0 \gtrsim 120$, the first-order transition appears for $\kappa_{\rm p}/\kappa_{\rm d}=3$ (see Fig.~\ref{fig:pd0}(a)).
With further increases in $H_0$, the coexistence region widens.
At a greater value of $\kappa_{\rm p}/\kappa_{\rm d}$, protein binding occurs 
at lower and higher values of $\mu$ for the buds and remaining region,
respectively, and the first-order transition starts at lower $H_0$ (see Figs.~\ref{fig:pd0}(b) and (c)).
For $H_0R_0\gg 1$, the large spherical component of the vesicle can be approximated as a flat membrane;
thus, $\mu$ for $\phi_{\rm L}=0.5$ is well represented by Eq.~(\ref{eq:phihalf}) for $H=0$
(see the overlay of dashed lines and solid gray lines in Fig.~\ref{fig:pd0}).
More precisely, the effect of the nonzero curvature is estimated as $-4\kappa_{\rm p}H_0a_{\rm p}/R_0$ from the second energy term in Eq.~(\ref{eq:phi}); thus, the deviation is proportional to $\kappa_{\rm p}$ and $H_0$ and is less than $0.1$ in  Fig.~\ref{fig:pd0}.
Moreover, the sigmoidal shapes of $\phi_{\rm L}$ shown in Fig.~\ref{fig:v9phi}
are well represented by 
\begin{equation} \label{eq:phisig}
\phi_{\rm L} = \frac{1}{1+\exp(-\frac{\mu-\mu_{\rm half}}{k_{\rm B}T})}.
\end{equation}
The deviation from Eq.~(\ref{eq:phisig}) is less than $10^{-5}$ for the data shown in Fig.~\ref{fig:v9phi}.
In contrast, the shape of $\phi_{\rm bud}$ can be largely modified.

Figure~\ref{fig:v} shows the dependence on the reduced volume $V^*$.
For smaller $V^*$,  $n_{\rm bud}$ increases to hold the excess area,
although no other notable differences are obtained for the vesicles with many buds.
The chemical potential  $\mu$ necessary for the transition to form many buds remains unchanged (see the dashed line in Fig.~\ref{fig:v}(b)).
In contrast,  $\mu$ required for the transition between the prolate to a single bud 
increases with increasing $V^*$ (see the lower two solid lines in Fig.~\ref{fig:v}(b) and the inset of Fig.~\ref{fig:v}(a)).
For $H_0R_0=200$, the prolate vesicle transforms into many buds without forming  a small number of buds,
so that $\mu$ for the transition is independent of $V^*$ (see the upper solid line in Fig.~\ref{fig:v}(b)).

Figure~\ref{fig:phi_dp} shows the dependence of $\phi_{\rm bud}$  on the area ratio $a_{\rm p}/R_0^2$.
The maximum number of bound proteins is $4\pi R_0^2/a_{\rm p}$.
For smaller $a_{\rm p}$ (or a larger vesicle),
 buds develop more gently with increasing $\mu$,
since more proteins bind to the vesicle and generate greater mixing entropy.
In contrast,  $\phi_{\rm L}$ is shifted to $2\kappa_{\rm p}H_0^2a_{\rm p}$,
and the sigmoid function of $\mu$ is maintained (data not shown).

Finally, we examine the validity of the chosen vesicle shapes.
First, the formation of a spherical bud on a prolate vesicle is considered.
We calculated the free energy of a prolate vesicle with a single bud for the conditions used in Fig.~\ref{fig:v}
and confirmed the energy is always higher than the spherical vesicle with a single bud or many buds.
Next, two sizes of the buds are allowed at  $H_0R_0=150$ (the other condition is the same in Figs.~\ref{fig:v9phi}--\ref{fig:v9bud}).
At the first-order transition point, the large buds of $R_{\rm min}/R_{\rm bud}=0.0365$ and $n_{\rm bud}=5$ 
and the small buds of $R_{\rm min}/R_{\rm bud}=0.891$ and $n_{\rm bud}= 2733$ coexist.
This large size of buds with a number of $2$, $3$, or $4$ and many small buds are prepared as initial states, 
and then we performed the energy minimization for $-1.8\le \mu/k_{\rm B}T\le -1.2$.
These two bud radii relax into the same value (the small or large sizes depending on the relaxation method and condition).
Thus, the assumption that the buds have the same radius is reasonable.

\subsubsection{Gaussian curvature}\label{sec:results2}

In this subsection, we examine the effects of the quantities related to the integral of the Gaussian curvature.
First, the effects of the modulation of the saddle-splay modulus $\bar{\kappa}$ are discussed.
When the protein does not modify $\bar{\kappa}$ ($\bar{\kappa}_{\rm p} = \bar{\kappa}_{\rm d}$),
the coexistence region is reduced at $\mu<0$ and begins at higher $H_0$, as shown in Fig.~\ref{fig:pdg}.
Note that the phase diagrams shown in Fig.~\ref{fig:pd0}
can represent the data for $\bar{\kappa}_{\rm p} = \bar{\kappa}_{\rm d}$
through rescaling with $\kappa_{\rm p}'= (\kappa_{\rm p}+\kappa_{\rm d})/2$, $H_0'=(\kappa_{\rm p}/\kappa_{\rm p}')H_0$, 
and $\mu'=\mu + 2\kappa_{\rm p}H_0^2a_{\rm p}(\kappa_{\rm p}/\kappa_{\rm p}' -1)$.
Thus, the phase behavior is qualitatively unchanged by the choice of the  $\bar{\kappa}$ dependence.

Next, the dependence on protein binding in the bud-neck region is considered.
When the necks have the same protein density as the large spherical component ($s_{\rm neck}=1$),
the boundary of the coexistence region slightly shifts to the right, as shown in Figs.~\ref{fig:phi_sa}(a)--(c),
and  $\phi_{\rm L}$ slightly decreases around the inflection point ($\phi_{\rm L}=0.5$), 
as shown in Fig.~\ref{fig:phi_sa}(d).
Therefore, the effects of $s_{\rm neck}$ are so small as to be negligible.

\begin{figure}[]
\includegraphics[]{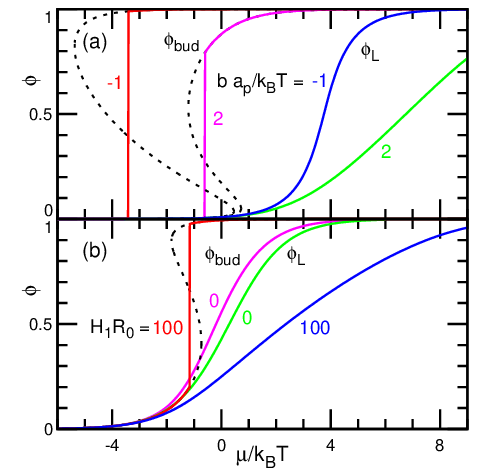}
\caption{
Dependence on protein pairwise interactions at $\kappa_{\rm p}/\kappa_{\rm d}=3$ and $V^*=0.9$.
(a) $b a_{\rm p}/k_{\rm B}T=-1$ and $2$ at $H_0R_0=200$ and $H_1=0$. (b) $H_1R_0=0$ and $100$ at $H_0R_0=50$ and $b=0$.
The solid lines represent thermal equilibrium states.
The two left and two right lines correspond to $\phi_{\rm bud}$ and $\phi_{\rm L}$, respectively.
The dashed lines represent metastable and free-energy-barrier states.
}
\label{fig:phi_bc}
\end{figure}

\begin{figure}[]
\includegraphics[]{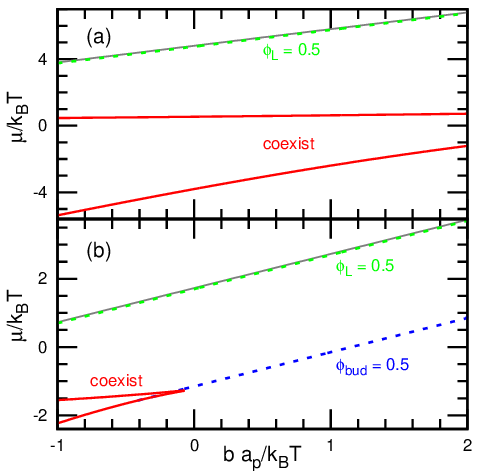}
\caption{
Phase diagrams of $b$ vs. $\mu$ for (a)  $H_0R_0=200$ and (b)  $H_0R_0=120$ at $V^*=0.9$, $\kappa_{\rm p}/\kappa_{\rm d}=3$, and $H_1=0$.
Two states with a small and large number of buds coexist in
the region between the two solid lines.
The dashed lines represent the states with $\phi_{\rm bud}=0.5$ or $\phi_{\rm L}=0.5$.
The solid gray lines are given by Eq.~(\ref{eq:phihalf}) and well overlay the data for $\phi_{\rm L}=0.5$.
}
\label{fig:pd_b}
\end{figure}

\begin{figure}[]
\includegraphics[]{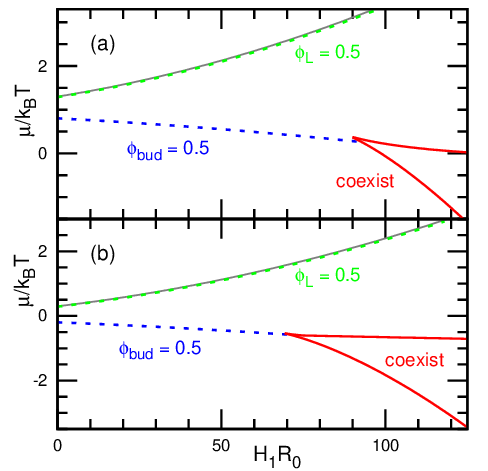}
\caption{
Phase diagram of $H_1$ vs. $\mu$ for (a) $b a_{\rm p}/k_{\rm B}T=1$ and (b) $b=0$ at $H_0R_0=50$, $\kappa_{\rm p}/\kappa_{\rm d}=3$, and $V^*=0.9$.
Two states with a small and large number of buds coexist in
the region between the two solid lines.
The upper and lower dashed lines represent the states with $\phi_{\rm L}=0.5$ and $\phi_{\rm bud}=0.5$, respectively.
The solid gray lines are given by Eq.~(\ref{eq:phihalf}) and well overlay the data for $\phi_{\rm L}=0.5$.
}
\label{fig:pd_d}
\end{figure}

\subsubsection{Effects of inter-protein interactions}\label{sec:results3}

Next, we consider the inter-protein interactions.
When $b$ is negative or positive, protein binding is more or less promoted at large $\phi$,
such that the coexistence region widens or shrinks, respectively (see Figs.~\ref{fig:phi_bc}(a) and \ref{fig:pd_b}).
Note that the membrane can be separated into regions with large and small $\phi$ within the buds or the main spherical component at $b < -2 k_{\rm B}T/a_{\rm p}$, in which $F_{\rm L}$ and $F_{\rm bud}$ can have double minima. However, because a homogeneous distribution is assumed for each component,
 such a phase-separation condition is not considered in this study.

For a nonzero value of $H_1$,
the spontaneous curvature increases from $H_0$ to $H_0+H_1$ as $\phi$ increases from $0$ to $1$;
thus, the coexistence region can appear with increasing $H_1$ (see Figs.~\ref{fig:phi_bc}(b) and \ref{fig:pd_d}).

The chemical potential $\mu$ for $\phi_{\rm L}=0.5$ continues to satisfy Eq.~(\ref{eq:phihalf}), as shown in Figs.~\ref{fig:pd_b} and \ref{fig:pd_d}.
However, 
the sigmoidal shape of $\phi_{\rm L}$ is  modified,
because $b$ and $H_1$ have greater influence at larger $\phi$ (see Fig.~\ref{fig:phi_bc}).
Therefore, when a significant deviation of $\phi_{\rm L}$ from the sigmoid function of  Eq.~(\ref{eq:phisig}) is observed,
the inter-protein interactions should be considered.

\begin{figure}[]
\includegraphics[]{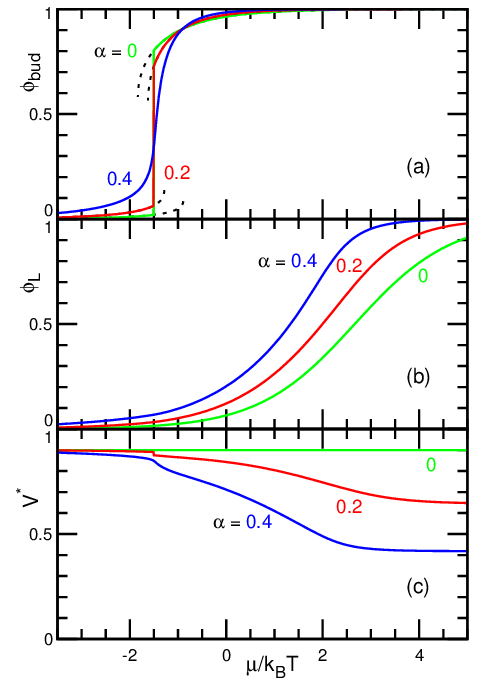}
\caption{
Dependence of (a) $\phi_{\rm bud}$, (b) $\phi_{\rm bud}$, and (c) $V^*$ on protein-insertion area ratio  $\alpha$ at $H_0R_0=150$, $\kappa_{\rm p}/\kappa_{\rm d}=3$, initial reduced volume  $V_{\rm int}^*=0.9$, and $b=H_1=0$.
The green, red, and blue lines represent the data for  $\alpha=0$, $0.2$, and $0.4$, respectively.
The dashed lines in (a) represent metastable states.
}
\label{fig:phi_al}
\end{figure}

\section{Area change due to protein insertion}\label{sec:area}

In this section, the changes in the membrane area induced by protein binding are considered.
When the proteins have a transmembrane domain, as depicted on the right side of Fig.~\ref{fig:cart}(a),
the membrane area is increased by the protein insertion.
Here, the area ratio of the transmembrane domain $\alpha \in [0,0.6]$ is considered.
Generally, lipid molecules are required to stabilize  the membrane proteins,
such that lipids remain between the proteins even under the most densely packed conditions.
Although the vesicle area is changed by this behavior, the lipid membrane area remains
at the initial value of $A_{\rm lp}= 4\pi R_0^2$. Thus,
\begin{equation} \label{eq:Ra}
R_0^2 = R_{\rm L}^2(1-\alpha\phi_{\rm L}) + \frac{A_{\rm bud}}{4\pi}(1-\alpha\phi_{\rm bud}).
\end{equation}
The vesicle volume is also fixed.
Hence, Eqs.~(\ref{eq:R0}) and (\ref{eq:VR}) are modified as
\begin{eqnarray} \label{eq:Abuda}
\frac{A_{\rm bud}}{4\pi} &=& \frac{ R_0^2[1 - r^2(1-\alpha\phi_{\rm L})]}{1-\alpha\phi_{\rm bud}}, \\ 
\frac{R_0}{R_{\rm bud}} &=& \frac{1-r^2(1-\alpha\phi_{\rm L})}{(1-\alpha\phi_{\rm bud})(V^*_{\rm int} - r^3)  },  \label{eq:Rbuda}
\end{eqnarray}
where $V^*_{\rm int}$ is the reduced volume of the initial vesicle (no protein binding).

The free energy is given by
\begin{eqnarray} 
\frac{F_{\rm L}}{4\pi} &=&  \label{eq:FL1a}
\bar{\kappa}_{\rm d} + 2(\kappa_{\rm dif}\phi_{\rm L}+\kappa_{\rm d}) -4\kappa_{\rm p}H_0R_0r\phi_{\rm L} \\ \nonumber
&& + R_0^2r^2(\sigma_{\rm p}\phi_{\rm L} + b \phi_{\rm L}^2) \\ \nonumber
&&+  \frac{(1-\alpha\phi_{\rm L}) k_{\rm B}TR_0^2 r^2}{(1-\alpha)a_{\rm p}}[\phi_{\rm L} \ln(\phi_{\rm L}) \\ \nonumber
&&+  (1-\phi_{\rm L}) \ln(1-\phi_{\rm L}) ],    \\ \nonumber
\frac{F_{\rm bud}}{4\pi} &=& n_{\rm bud} 
\Big\{  2(\kappa_{\rm dif}\phi_{\rm bud}+\kappa_{\rm d})
 - 4 \kappa_{\rm p}H_0R_{\rm bud}\phi_{\rm bud}  \\ \nonumber
&& + {R_{\rm bud}}^2(\sigma_{\rm p}\phi_{\rm bud} + b \phi_{\rm bud}^2 ) \\ \nonumber
&&  + \frac{(1-\alpha\phi_{\rm bud})k_{\rm B}T {R_{\rm bud}}^2}{(1-\alpha)a_{\rm p}}[\phi_{\rm bud} \ln(\phi_{\rm bud})  \\ 
&& +  (1-\phi_{\rm bud}) \ln(1-\phi_{\rm bud}) ]   \Big\} \label{eq:FB1a}  \\ \nonumber
 &=&  
 2(\kappa_{\rm dif}\phi_{\rm bud}+\kappa_{\rm d})\frac{[1-r^2(1-\alpha\phi_{\rm L})]^3}{(1-\alpha\phi_{\rm bud})^3(V^* - r^3)^2}  \\ \nonumber
&& - 4 \kappa_{\rm p}H_0R_0 \phi_{\rm bud}\frac{[1-r^2(1-\alpha\phi_{\rm L})]^2}{(1-\alpha\phi_{\rm bud})^2(V^* - r^3)}   \\ \nonumber
&& + \frac{R_0^2[1 - r^2(1-\alpha\phi_{\rm L})]}{1-\alpha\phi_{\rm bud}}  (\sigma_{\rm p}\phi_{\rm bud} + b \phi_{\rm bud}^2) \\ \nonumber
&&  + \frac{[1-r^2(1-\alpha\phi_{\rm L})] k_{\rm B}TR_0^2}{(1-\alpha)a_{\rm p}}[\phi_{\rm bud} \ln(\phi_{\rm bud})  \\ 
&& +  (1-\phi_{\rm bud}) \ln(1-\phi_{\rm bud}) ]. \label{eq:FB2a}
\end{eqnarray}
The coefficients of the entropy terms are modified by the factors $(1-\alpha\phi_{\rm L})/(1-\alpha)$
and $(1-\alpha\phi_{\rm bud})/(1-\alpha)$ to count the number of binding sites under the area change.

\begin{figure}[]
\includegraphics[]{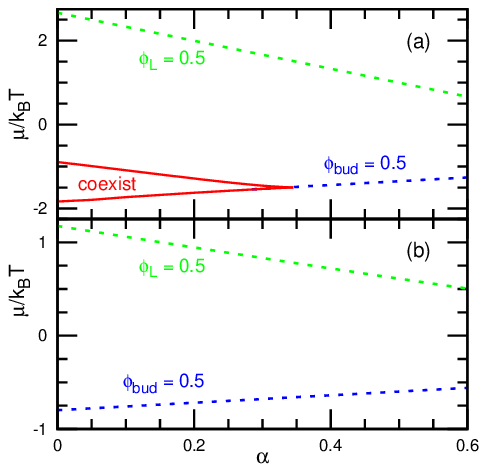}
\caption{
Phase diagram of $\alpha$ vs. $\mu$ for (a) $H_0R_0=150$  and (b) $H_0R_0=100$ at $\kappa_{\rm p}/\kappa_{\rm d}=3$, $V^*=0.9$, and $b=H_1=0$.
Two states with a small and large number of buds coexist in
the region between the two solid lines.
The upper and lower dashed lines represent the states with $\phi_{\rm L}=0.5$ and $\phi_{\rm bud}=0.5$, respectively.
}
\label{fig:pd_al}
\end{figure}

Figures~\ref{fig:phi_al} and \ref{fig:pd_al} show the dependences of various properties on $\alpha$.
With increasing $\alpha$, the coexistence region shrinks (see Figs.~\ref{fig:phi_al}(a) and \ref{fig:pd_al}).
Further, $V^*$ decreases with increasing $\phi_{\rm L}$ at $\alpha>0$, 
since the surface area is increased by the protein insertion (see Fig.~\ref{fig:phi_al}(c)).
Accordingly, the sigmoidal shape of  $\phi_{\rm L}$ is slightly modified (see Fig.~\ref{fig:phi_al}(b)).

\section{Effects of area-difference elasticity}\label{sec:ade}

In this section, the effects of the area-difference elasticity (ADE) energy on the vesicle budding are considered.
In lipid membranes,
the flip--flop (traverse motion between monolayers) of lipids
is very slow; the relaxation time of the phospholipids is several hours to days~\cite{korn71}.
In contrast, amphiphilic molecules with small hydrophilic head groups such as 
cholesterols and diacylglycerols exhibit much faster flip--flop (less than a minute)~\cite{cont09,hami03,stec02}.
In living cells, proteins transport lipids in addition;
flippases or floppases pump specific lipids in one direction (flip or flop, respectively) using ATP hydrolysis, 
yielding an asymmetric lipid distribution, whereas
scramblases translocate lipids in both directions and relax the bilayer into a thermal-equilibrium lipid distribution~\cite{cont09}. 
Thus, the number of lipids in each monolayer can be fixed on time scales of typical {\it in vitro} experiments,
although it can relax with the addition of cholesterols~\cite{bruc09,miet19},  ultra-long-chain fatty acids ~\cite{kawa20}, and scramblases.
Hence, area difference $\Delta A_0=(N_{\rm {out}}-N_{\rm {in}})a_0$
preferred by lipids can be different from the area difference $\Delta A$ of the vesicle,
where $N_{\rm {out}}$ and $N_{\rm {in}}$
are the numbers of lipids in the outer and inner monolayers, respectively,
and $a_0$ is the area per lipid.
In the ADE model~\cite{seif97,svet09,svet89}, 
the energy of this mismatch $\Delta A-\Delta A_0$ is taken into account by a harmonic potential:
\begin{equation}
F_{\rm {ade}} =  \frac{\pi k_{\rm {ade}}}{2Ah^2}(\Delta A - \Delta A_0)^2
=  \frac{k_{\rm {ade}}}{2}(m - m_0)^2,
\label{eq:ade0}
\end{equation}
with the averaged curvature $m= (1/2R_0)\oint (C_1+C_2) dA$.
For typical lipid membranes,
 $k_{\rm {ade}} \simeq \kappa$ has been reported~\cite{saka12} 
($q = \pi k_{\rm {ade}}/\kappa$ in the notation in Ref.~\cite{saka12}).

\begin{figure}[]
\includegraphics[]{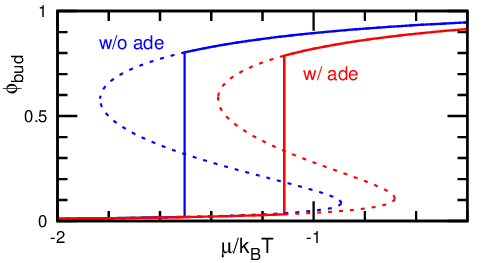}
\caption{
Effects of the ADE energy on
 $\phi_{\rm bud}$ at $H_0R_0=150$, $\kappa_{\rm p}/\kappa_{\rm d}=3$, $V_{\rm int}^*=0.9$, and $b=H_1=0$,
The red and blue colors represent the data with and without the ADE energy, respectively.
}
\label{fig:phi_ade}
\end{figure}

\begin{figure}[]
\includegraphics[]{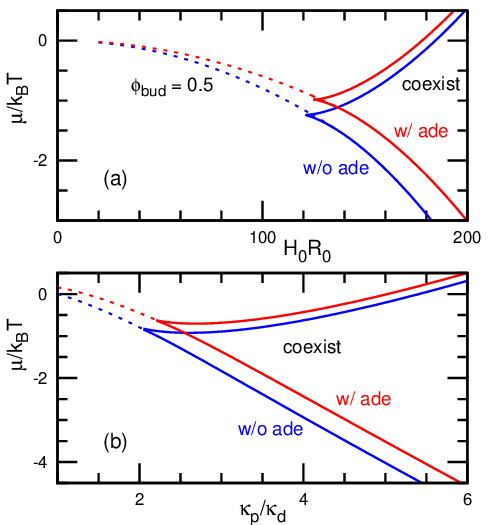}
\caption{
Effects of the ADE energy on phase diagrams at $V^*=0.9$ and $b=H_1=0$.
(a) Phase diagram of $H_0R_0$ vs. $\mu$ for $\kappa_{\rm p}/\kappa_{\rm d}=3$.
(b) Phase diagram of $\kappa_{\rm p}/\kappa_{\rm d}$ vs. $\mu$ for $H_0R_0=150$.
Two states with a small and large number of buds coexist in
the region between the two solid lines.
The dashed lines represent the states with $\phi_{\rm bud}=0.5$.
The upper (red) and lower (blue) lines represent the data with and without the ADE energy, respectively.
}
\label{fig:pd_ade}
\end{figure}

The area differences of the budded and prolate vesicles are given by
\begin{eqnarray} \label{eq:adb}
\frac{m_{\rm ves}}{4\pi} &=& r + n_{\rm bud}\frac{R_{\rm bud}}{R_0}, \\
\frac{m_{\rm pro}}{4\pi} &=& \frac{R_{\rm pro}}{R_0} + \frac{L_{\rm pro}}{4R_0},
\label{eq:adp}
\end{eqnarray}
respectively.
Here, the case of no flip--flop and $m_0=m_{\rm pro}$ is considered; that is,
 the initial prolate vesicle has no ADE penalty.
This corresponds to an experimental condition on a time scale spanning minutes to a few hours,
in the absence of sterols and proteins promoting flip--flop or forming membrane pores.

Figures~\ref{fig:phi_ade} and \ref{fig:pd_ade} show the effects of the ADE energy.
Here, $\mu$ becomes slightly larger to overcome  the ADE energy,
but the amount is less than $k_{\rm B}T$.
The influence on $\phi_{\rm L}$ is negligibly small (data not shown).
Thus, the effects can only be detected if the other parameters are well determined.
Otherwise, the ADE energy can be neglected.

In the above analysis, it is assumed that the bound proteins do not change the preferred area difference $m_0$ between the two monolayers.
However, some proteins may modify $m_0$ through the domain insertion into the membrane.
When protein binding significantly increases $m_0$, 
the buds form at smaller $\mu$ and the phase boundaries shown in Fig.~\ref{fig:pd_ade} 
are likely shifted downward.

\section{Discussion}\label{sec:dis}

In this section, we discuss the experimental conditions for vesicle budding.
Note that $\mu$ can be controlled by the buffer protein concentration $\rho$.
For a dilute solution, it is given by $\mu=\mu_0 + k_{\rm B}T\ln(\rho)$.
Through screening of electrostatic interactions,
$\mu$ also depends on the ion concentration.

Some proteins have large hydrophilic domains
that can generate a repulsive interaction between them ($b>0$),
and protein-density-dependent spontaneous curvature ($H_1>0$).
The dependence of the spontaneous curvature on the length of the disordered hydrophilic domains
has been investigated experimentally~\cite{busc15,zeno19}. 
The induced spontaneous curvature can be analytically represented
using the conformational entropy of the chains~\cite{hier96,bick06,wu13} and the excluded volume~\cite{tozz19}.
Although  a linear dependence on the protein density was considered in this work,
it can easily be extended to include a nonlinear dependence for a specific protein.

Many proteins have hydrophobic segments to insert the membrane.
 G-protein-coupled receptors and ion channels have wide transmembrane domains,
some of which have been reported to sense and generate membrane curvature~\cite{aimo14,rosh17}.
For these proteins, the area change ($\alpha$) due to protein binding should be considered.
However, when the area change is small ($\alpha \ll 1$),
it can be neglected, because the $\alpha$ dependence is not sensitive.
For example, 
the diameters of $\alpha$HL  are $10$\, nm and $1.2$\,nm at the head and transmembrane regions, respectively
($\alpha \simeq 0.01$)~\cite{andr07}.
Thus, the $\alpha$HL binding can be approximated as $\alpha=0$.
The area change due to shallow insertion of domains such as amphipathic $\alpha$-helix is also likely negligible.
Some transmembrane proteins require the help of other proteins for  membrane insertion.
In the experiments of such proteins, the total number of proteins is rather fixed in the membrane.
To investigate them, 
the present model can be modified to control the total number of proteins instead of the chemical potential.

Protein assembly can be induced by the inter-protein attraction, as seen in the clathrin binding~\cite{avin15,sale15,hass17,frey20}.
To model this behavior, the line tension of the phase boundary should be incorporated into the model, 
and a constant protein density is used in the assembled domains.
Moreover, the bound proteins may exhibit a conformational change or form a complex of several proteins  
through membrane--protein interactions, as depicted on the right side of Fig.~\ref{fig:cart}(a).
When different binding conformations coexist in significant amounts on the membrane in thermal equilibrium,
multiple bound states can be considered. The present model is easily extended to this scenario, 
although many more parameters (the free-energy parameters for each state) 
must be determined.

Under some experimental conditions,
the buds may be pinched off and form separated vesicles.
Moreover, the membrane rupture may be caused by the high surface tension induced by protein binding.
These dynamics can be taken into account by setting a threshold of the surface tension or the bud radius,
which depends on the membrane composition and proteins.

\section{Summary}\label{sec:sum}

We have studied vesicle budding using mean-field theory.
First, we presented a formula for  protein binding with an arbitrary curvature
and clarified the preferred curvatures for curvature sensing and generation.
The generation curvature is lower than the sensing curvature, because the proteins are required to bend the membrane.
Then, we presented the free energy of the budded vesicle, in which the protein binding can modify 
the spontaneous curvature, bending rigidity,  saddle-splay modulus of the membrane.
Moreover, inter-protein interactions are included as direct interaction and
protein-density dependent spontaneous curvature.
Furthermore, the changes in the membrane area due to the protein insertion and ADE energy
are taken into account.

With increasing binding chemical potential $\mu$,
a prolate vesicle transforms into a budded vesicle. Subsequently,
the number of buds increases and the bud radius is saturated to that for the curvature generation.
The surface tension $\sigma_{\rm d}$ increases linearly to $\mu$ after bud formation.
Further, when high spontaneous curvature $H_0$ and/or large bending rigidity $\kappa_{\rm p}$ are induced by the protein binding,
a first-order shape transition occurs between a vesicle with a small number of large buds
and that with a large number of small buds.
These two states  coexist in a wider $\mu$ range for higher $H_0$ or larger $\kappa_{\rm p}$.
The attractive interaction between bound proteins and the spontaneous-curvature increase due to the protein contact
promote budding and generate a wider coexistence region.
In contrast, the area increase due to protein insertion reduces the coexistence region.
When the preferred area difference between two monolayers is fixed at the initial prolate,
the budding requires a slightly larger $\mu$ owing to the ADE energy.

In this study, we approximated that the budded vesicle consists of spheres.
With the help of this simple geometry, the free-energy minimum could be  determined so easily
 that additional terms could be considered.
Hence, the effects of protein--membrane and inter-protein interactions
were examined.
However, the proposed model can be extended to more complicated geometries.
Under low  $\sigma_{\rm d}$,
a flat membrane can exhibit 
a separated/corrugated (SC) phase, in which the bound proteins form hexagonally ordered bowl-shaped domains,
even in the absence of direct inter-protein attractive interactions~\cite{gout21}.
Hence, this SC phase may appear on a vesicle before budding or on a vesicle with a few buds.
These bowl-shaped domains should be modeled into a simplified geometry to account for the SC phase.

Moreover, the present strategy can be used to examine other types of interactions, as discussed in Sec.~\ref{sec:dis},
and extended to include other geometries.
For example, BAR superfamily proteins can induce cylindrical membrane tubes from liposomes~\cite{mcma05,suet14,joha15,itoh06,mim12a}.
Recently, the formation of a nematic order coupled with the bending energy was investigated for a membrane with fixed shapes using a mean-field theory~\cite{tozz21}. 
The stable shape of these membrane tubes that protrude from a vesicle
is an interesting problem for further study.
As discussed above, the model presented herein is a useful tool for investigating many problems related to vesicle deformation and can be further extended in various directions.

\begin{acknowledgments}
We thank M. Yanagisawa (Univ. Tokyo) for stimulating discussions.
This work was supported by JSPS KAKENHI Grant Number JP21K03481.
\end{acknowledgments}


\begin{thebibliography}{78}%
\makeatletter
\providecommand \@ifxundefined [1]{%
 \@ifx{#1\undefined}
}%
\providecommand \@ifnum [1]{%
 \ifnum #1\expandafter \@firstoftwo
 \else \expandafter \@secondoftwo
 \fi
}%
\providecommand \@ifx [1]{%
 \ifx #1\expandafter \@firstoftwo
 \else \expandafter \@secondoftwo
 \fi
}%
\providecommand \natexlab [1]{#1}%
\providecommand \enquote  [1]{``#1''}%
\providecommand \bibnamefont  [1]{#1}%
\providecommand \bibfnamefont [1]{#1}%
\providecommand \citenamefont [1]{#1}%
\providecommand \href@noop [0]{\@secondoftwo}%
\providecommand \href [0]{\begingroup \@sanitize@url \@href}%
\providecommand \@href[1]{\@@startlink{#1}\@@href}%
\providecommand \@@href[1]{\endgroup#1\@@endlink}%
\providecommand \@sanitize@url [0]{\catcode `\\12\catcode `\$12\catcode
  `\&12\catcode `\#12\catcode `\^12\catcode `\_12\catcode `\%12\relax}%
\providecommand \@@startlink[1]{}%
\providecommand \@@endlink[0]{}%
\providecommand \url  [0]{\begingroup\@sanitize@url \@url }%
\providecommand \@url [1]{\endgroup\@href {#1}{\urlprefix }}%
\providecommand \urlprefix  [0]{URL }%
\providecommand \Eprint [0]{\href }%
\providecommand \doibase [0]{http://dx.doi.org/}%
\providecommand \selectlanguage [0]{\@gobble}%
\providecommand \bibinfo  [0]{\@secondoftwo}%
\providecommand \bibfield  [0]{\@secondoftwo}%
\providecommand \translation [1]{[#1]}%
\providecommand \BibitemOpen [0]{}%
\providecommand \bibitemStop [0]{}%
\providecommand \bibitemNoStop [0]{.\EOS\space}%
\providecommand \EOS [0]{\spacefactor3000\relax}%
\providecommand \BibitemShut  [1]{\csname bibitem#1\endcsname}%
\let\auto@bib@innerbib\@empty
\bibitem [{\citenamefont {McMahon}\ and\ \citenamefont
  {Gallop}(2005)}]{mcma05}%
  \BibitemOpen
  \bibfield  {author} {\bibinfo {author} {\bibfnamefont {H.~T.}\ \bibnamefont
  {McMahon}}\ and\ \bibinfo {author} {\bibfnamefont {J.~L.}\ \bibnamefont
  {Gallop}},\ }\href@noop {} {\bibfield  {journal} {\bibinfo  {journal}
  {Nature}\ }\textbf {\bibinfo {volume} {438}},\ \bibinfo {pages} {590}
  (\bibinfo {year} {2005})}\BibitemShut {NoStop}%
\bibitem [{\citenamefont {Suetsugu}\ \emph {et~al.}(2014)\citenamefont
  {Suetsugu}, \citenamefont {Kurisu},\ and\ \citenamefont {Takenawa}}]{suet14}%
  \BibitemOpen
  \bibfield  {author} {\bibinfo {author} {\bibfnamefont {S.}~\bibnamefont
  {Suetsugu}}, \bibinfo {author} {\bibfnamefont {S.}~\bibnamefont {Kurisu}}, \
  and\ \bibinfo {author} {\bibfnamefont {T.}~\bibnamefont {Takenawa}},\
  }\href@noop {} {\bibfield  {journal} {\bibinfo  {journal} {Physiol. Rev.}\
  }\textbf {\bibinfo {volume} {94}},\ \bibinfo {pages} {1219} (\bibinfo {year}
  {2014})}\BibitemShut {NoStop}%
\bibitem [{\citenamefont {Johannes}\ \emph {et~al.}(2015)\citenamefont
  {Johannes}, \citenamefont {Parton}, \citenamefont {Bassereau},\ and\
  \citenamefont {Mayor}}]{joha15}%
  \BibitemOpen
  \bibfield  {author} {\bibinfo {author} {\bibfnamefont {L.}~\bibnamefont
  {Johannes}}, \bibinfo {author} {\bibfnamefont {R.~G.}\ \bibnamefont
  {Parton}}, \bibinfo {author} {\bibfnamefont {P.}~\bibnamefont {Bassereau}}, \
  and\ \bibinfo {author} {\bibfnamefont {S.}~\bibnamefont {Mayor}},\
  }\href@noop {} {\bibfield  {journal} {\bibinfo  {journal} {Nat. Rev. Mol.
  Cell. Biol.}\ }\textbf {\bibinfo {volume} {16}},\ \bibinfo {pages} {311}
  (\bibinfo {year} {2015})}\BibitemShut {NoStop}%
\bibitem [{\citenamefont {Brandizzi}\ and\ \citenamefont
  {Barlowe}(2013)}]{bran13}%
  \BibitemOpen
  \bibfield  {author} {\bibinfo {author} {\bibfnamefont {F.}~\bibnamefont
  {Brandizzi}}\ and\ \bibinfo {author} {\bibfnamefont {C.}~\bibnamefont
  {Barlowe}},\ }\href@noop {} {\bibfield  {journal} {\bibinfo  {journal} {Nat.
  Rev. Mol. Cell Biol.}\ }\textbf {\bibinfo {volume} {14}},\ \bibinfo {pages}
  {382} (\bibinfo {year} {2013})}\BibitemShut {NoStop}%
\bibitem [{\citenamefont {Hurley}\ \emph {et~al.}(2010)\citenamefont {Hurley},
  \citenamefont {Boura}, \citenamefont {Carlson},\ and\ \citenamefont
  {R{\'o}{\.{z}}ycki}}]{hurl10}%
  \BibitemOpen
  \bibfield  {author} {\bibinfo {author} {\bibfnamefont {J.~H.}\ \bibnamefont
  {Hurley}}, \bibinfo {author} {\bibfnamefont {E.}~\bibnamefont {Boura}},
  \bibinfo {author} {\bibfnamefont {L.-A.}\ \bibnamefont {Carlson}}, \ and\
  \bibinfo {author} {\bibfnamefont {B.}~\bibnamefont {R{\'o}{\.{z}}ycki}},\
  }\href@noop {} {\bibfield  {journal} {\bibinfo  {journal} {Cell}\ }\textbf
  {\bibinfo {volume} {143}},\ \bibinfo {pages} {875} (\bibinfo {year}
  {2010})}\BibitemShut {NoStop}%
\bibitem [{\citenamefont {McMahon}\ and\ \citenamefont
  {Boucrot}(2011)}]{mcma11}%
  \BibitemOpen
  \bibfield  {author} {\bibinfo {author} {\bibfnamefont {H.~T.}\ \bibnamefont
  {McMahon}}\ and\ \bibinfo {author} {\bibfnamefont {E.}~\bibnamefont
  {Boucrot}},\ }\href@noop {} {\bibfield  {journal} {\bibinfo  {journal} {Nat.
  Rev. Mol. Cell. Biol.}\ }\textbf {\bibinfo {volume} {12}},\ \bibinfo {pages}
  {517} (\bibinfo {year} {2011})}\BibitemShut {NoStop}%
\bibitem [{\citenamefont {Schmid}\ and\ \citenamefont {Frolov}(2011)}]{schm11}%
  \BibitemOpen
  \bibfield  {author} {\bibinfo {author} {\bibfnamefont {S.~L.}\ \bibnamefont
  {Schmid}}\ and\ \bibinfo {author} {\bibfnamefont {V.~A.}\ \bibnamefont
  {Frolov}},\ }\href@noop {} {\bibfield  {journal} {\bibinfo  {journal} {Annu.
  Rev. Cell Dev. Biol.}\ }\textbf {\bibinfo {volume} {27}},\ \bibinfo {pages}
  {79} (\bibinfo {year} {2011})}\BibitemShut {NoStop}%
\bibitem [{\citenamefont {Kaksonen}\ and\ \citenamefont {Roux}(2018)}]{kaks18}%
  \BibitemOpen
  \bibfield  {author} {\bibinfo {author} {\bibfnamefont {M.}~\bibnamefont
  {Kaksonen}}\ and\ \bibinfo {author} {\bibfnamefont {A.}~\bibnamefont
  {Roux}},\ }\href@noop {} {\bibfield  {journal} {\bibinfo  {journal} {Nat.
  Rev. Mol. Cell Biol.}\ }\textbf {\bibinfo {volume} {19}},\ \bibinfo {pages}
  {313} (\bibinfo {year} {2018})}\BibitemShut {NoStop}%
\bibitem [{\citenamefont {Avinoam}\ \emph {et~al.}(2015)\citenamefont
  {Avinoam}, \citenamefont {Schorb}, \citenamefont {Beese}, \citenamefont
  {Briggs},\ and\ \citenamefont {Kaksonen}}]{avin15}%
  \BibitemOpen
  \bibfield  {author} {\bibinfo {author} {\bibfnamefont {O.}~\bibnamefont
  {Avinoam}}, \bibinfo {author} {\bibfnamefont {M.}~\bibnamefont {Schorb}},
  \bibinfo {author} {\bibfnamefont {C.~J.}\ \bibnamefont {Beese}}, \bibinfo
  {author} {\bibfnamefont {J.~A.~G.}\ \bibnamefont {Briggs}}, \ and\ \bibinfo
  {author} {\bibfnamefont {M.}~\bibnamefont {Kaksonen}},\ }\href@noop {}
  {\bibfield  {journal} {\bibinfo  {journal} {Science}\ }\textbf {\bibinfo
  {volume} {348}},\ \bibinfo {pages} {1369} (\bibinfo {year}
  {2015})}\BibitemShut {NoStop}%
\bibitem [{\citenamefont {Seifert}(1997)}]{seif97}%
  \BibitemOpen
  \bibfield  {author} {\bibinfo {author} {\bibfnamefont {U.}~\bibnamefont
  {Seifert}},\ }\href@noop {} {\bibfield  {journal} {\bibinfo  {journal} {Adv.\
  Phys.}\ }\textbf {\bibinfo {volume} {46}},\ \bibinfo {pages} {13} (\bibinfo
  {year} {1997})}\BibitemShut {NoStop}%
\bibitem [{\citenamefont {Svetina}(2009)}]{svet09}%
  \BibitemOpen
  \bibfield  {author} {\bibinfo {author} {\bibfnamefont {S.}~\bibnamefont
  {Svetina}},\ }\href@noop {} {\bibfield  {journal} {\bibinfo  {journal}
  {ChemPhysChem}\ }\textbf {\bibinfo {volume} {10}},\ \bibinfo {pages} {2769}
  (\bibinfo {year} {2009})}\BibitemShut {NoStop}%
\bibitem [{\citenamefont {Hotani}\ \emph {et~al.}(1999)\citenamefont {Hotani},
  \citenamefont {Nomura},\ and\ \citenamefont {Suzuki}}]{hota99}%
  \BibitemOpen
  \bibfield  {author} {\bibinfo {author} {\bibfnamefont {H.}~\bibnamefont
  {Hotani}}, \bibinfo {author} {\bibfnamefont {F.}~\bibnamefont {Nomura}}, \
  and\ \bibinfo {author} {\bibfnamefont {Y.}~\bibnamefont {Suzuki}},\
  }\href@noop {} {\bibfield  {journal} {\bibinfo  {journal} {Curr.\ Opin.\
  Coll.\ Interface\ Sci.}\ }\textbf {\bibinfo {volume} {4}},\ \bibinfo {pages}
  {358} (\bibinfo {year} {1999})}\BibitemShut {NoStop}%
\bibitem [{\citenamefont {Sakashita}\ \emph {et~al.}(2012)\citenamefont
  {Sakashita}, \citenamefont {Urakami}, \citenamefont {Ziherl},\ and\
  \citenamefont {Imai}}]{saka12}%
  \BibitemOpen
  \bibfield  {author} {\bibinfo {author} {\bibfnamefont {A.}~\bibnamefont
  {Sakashita}}, \bibinfo {author} {\bibfnamefont {N.}~\bibnamefont {Urakami}},
  \bibinfo {author} {\bibfnamefont {P.}~\bibnamefont {Ziherl}}, \ and\ \bibinfo
  {author} {\bibfnamefont {M.}~\bibnamefont {Imai}},\ }\href@noop {} {\bibfield
   {journal} {\bibinfo  {journal} {Soft\ Matter}\ }\textbf {\bibinfo {volume}
  {8}},\ \bibinfo {pages} {8569} (\bibinfo {year} {2012})}\BibitemShut
  {NoStop}%
\bibitem [{\citenamefont {Holl{\'o}}\ \emph {et~al.}(2021)\citenamefont
  {Holl{\'o}}, \citenamefont {Miele}, \citenamefont {Rossi},\ and\
  \citenamefont {Lagzi}}]{holl21}%
  \BibitemOpen
  \bibfield  {author} {\bibinfo {author} {\bibfnamefont {G.}~\bibnamefont
  {Holl{\'o}}}, \bibinfo {author} {\bibfnamefont {Y.}~\bibnamefont {Miele}},
  \bibinfo {author} {\bibfnamefont {F.}~\bibnamefont {Rossi}}, \ and\ \bibinfo
  {author} {\bibfnamefont {I.}~\bibnamefont {Lagzi}},\ }\href@noop {}
  {\bibfield  {journal} {\bibinfo  {journal} {Phys. Chem. Chem. Phys.}\
  }\textbf {\bibinfo {volume} {23}},\ \bibinfo {pages} {4262} (\bibinfo {year}
  {2021})}\BibitemShut {NoStop}%
\bibitem [{\citenamefont {Baumgart}\ \emph {et~al.}(2003)\citenamefont
  {Baumgart}, \citenamefont {Hess},\ and\ \citenamefont {Webb}}]{baum03}%
  \BibitemOpen
  \bibfield  {author} {\bibinfo {author} {\bibfnamefont {T.}~\bibnamefont
  {Baumgart}}, \bibinfo {author} {\bibfnamefont {S.~T.}\ \bibnamefont {Hess}},
  \ and\ \bibinfo {author} {\bibfnamefont {W.~W.}\ \bibnamefont {Webb}},\
  }\href@noop {} {\bibfield  {journal} {\bibinfo  {journal} {Nature}\ }\textbf
  {\bibinfo {volume} {425}},\ \bibinfo {pages} {821} (\bibinfo {year}
  {2003})}\BibitemShut {NoStop}%
\bibitem [{\citenamefont {Bacia}\ \emph {et~al.}(2005)\citenamefont {Bacia},
  \citenamefont {Schwille},\ and\ \citenamefont {Kurzchalia}}]{baci05}%
  \BibitemOpen
  \bibfield  {author} {\bibinfo {author} {\bibfnamefont {K.}~\bibnamefont
  {Bacia}}, \bibinfo {author} {\bibfnamefont {P.}~\bibnamefont {Schwille}}, \
  and\ \bibinfo {author} {\bibfnamefont {T.}~\bibnamefont {Kurzchalia}},\
  }\href@noop {} {\bibfield  {journal} {\bibinfo  {journal} {Proc.\ Natl.\
  Acad.\ Sci.\ USA}\ }\textbf {\bibinfo {volume} {102}},\ \bibinfo {pages}
  {3272} (\bibinfo {year} {2005})}\BibitemShut {NoStop}%
\bibitem [{\citenamefont {Yanagisawa}\ \emph {et~al.}(2008)\citenamefont
  {Yanagisawa}, \citenamefont {Imai},\ and\ \citenamefont
  {Taniguchi}}]{yana08}%
  \BibitemOpen
  \bibfield  {author} {\bibinfo {author} {\bibfnamefont {M.}~\bibnamefont
  {Yanagisawa}}, \bibinfo {author} {\bibfnamefont {M.}~\bibnamefont {Imai}}, \
  and\ \bibinfo {author} {\bibfnamefont {T.}~\bibnamefont {Taniguchi}},\
  }\href@noop {} {\bibfield  {journal} {\bibinfo  {journal} {Phys.\ Rev.\
  Lett.}\ }\textbf {\bibinfo {volume} {100}},\ \bibinfo {pages} {148102}
  (\bibinfo {year} {2008})}\BibitemShut {NoStop}%
\bibitem [{\citenamefont {Tsafrir}\ \emph {et~al.}(2001)\citenamefont
  {Tsafrir}, \citenamefont {Sagi}, \citenamefont {Arzi}, \citenamefont
  {Guedeau-Boudeville}, \citenamefont {Frette}, \citenamefont {Kandel},\ and\
  \citenamefont {Stavans}}]{tsaf01}%
  \BibitemOpen
  \bibfield  {author} {\bibinfo {author} {\bibfnamefont {I.}~\bibnamefont
  {Tsafrir}}, \bibinfo {author} {\bibfnamefont {D.}~\bibnamefont {Sagi}},
  \bibinfo {author} {\bibfnamefont {T.}~\bibnamefont {Arzi}}, \bibinfo {author}
  {\bibfnamefont {M.-A.}\ \bibnamefont {Guedeau-Boudeville}}, \bibinfo {author}
  {\bibfnamefont {V.}~\bibnamefont {Frette}}, \bibinfo {author} {\bibfnamefont
  {D.}~\bibnamefont {Kandel}}, \ and\ \bibinfo {author} {\bibfnamefont
  {J.}~\bibnamefont {Stavans}},\ }\href@noop {} {\bibfield  {journal} {\bibinfo
   {journal} {Phys. Rev. Lett.}\ }\textbf {\bibinfo {volume} {86}},\ \bibinfo
  {pages} {1138} (\bibinfo {year} {2001})}\BibitemShut {NoStop}%
\bibitem [{\citenamefont {Saleem}\ \emph {et~al.}(2015)\citenamefont {Saleem},
  \citenamefont {Morlot}, \citenamefont {Hohendahl}, \citenamefont {Manzi},
  \citenamefont {Lenz},\ and\ \citenamefont {Roux}}]{sale15}%
  \BibitemOpen
  \bibfield  {author} {\bibinfo {author} {\bibfnamefont {M.}~\bibnamefont
  {Saleem}}, \bibinfo {author} {\bibfnamefont {S.}~\bibnamefont {Morlot}},
  \bibinfo {author} {\bibfnamefont {A.}~\bibnamefont {Hohendahl}}, \bibinfo
  {author} {\bibfnamefont {J.}~\bibnamefont {Manzi}}, \bibinfo {author}
  {\bibfnamefont {M.}~\bibnamefont {Lenz}}, \ and\ \bibinfo {author}
  {\bibfnamefont {A.}~\bibnamefont {Roux}},\ }\href@noop {} {\bibfield
  {journal} {\bibinfo  {journal} {Nat. Commun.}\ }\textbf {\bibinfo {volume}
  {6}},\ \bibinfo {pages} {6249} (\bibinfo {year} {2015})}\BibitemShut
  {NoStop}%
\bibitem [{\citenamefont {Boye}\ \emph {et~al.}(2018)\citenamefont {Boye},
  \citenamefont {Jeppesen}, \citenamefont {Maeda}, \citenamefont {Pezeshkian},
  \citenamefont {Solovyeva}, \citenamefont {Nylandsted},\ and\ \citenamefont
  {Simonsen}}]{boye18}%
  \BibitemOpen
  \bibfield  {author} {\bibinfo {author} {\bibfnamefont {T.~L.}\ \bibnamefont
  {Boye}}, \bibinfo {author} {\bibfnamefont {J.~C.}\ \bibnamefont {Jeppesen}},
  \bibinfo {author} {\bibfnamefont {K.}~\bibnamefont {Maeda}}, \bibinfo
  {author} {\bibfnamefont {W.}~\bibnamefont {Pezeshkian}}, \bibinfo {author}
  {\bibfnamefont {V.}~\bibnamefont {Solovyeva}}, \bibinfo {author}
  {\bibfnamefont {J.}~\bibnamefont {Nylandsted}}, \ and\ \bibinfo {author}
  {\bibfnamefont {A.~C.}\ \bibnamefont {Simonsen}},\ }\href@noop {} {\bibfield
  {journal} {\bibinfo  {journal} {Sci. Rep.}\ }\textbf {\bibinfo {volume}
  {8}},\ \bibinfo {pages} {10309} (\bibinfo {year} {2018})}\BibitemShut
  {NoStop}%
\bibitem [{\citenamefont {Itoh}\ and\ \citenamefont {{De
  Camilli}}(2006)}]{itoh06}%
  \BibitemOpen
  \bibfield  {author} {\bibinfo {author} {\bibfnamefont {T.}~\bibnamefont
  {Itoh}}\ and\ \bibinfo {author} {\bibfnamefont {P.}~\bibnamefont {{De
  Camilli}}},\ }\href@noop {} {\bibfield  {journal} {\bibinfo  {journal}
  {Biochim.\ Biophys.\ Acta}\ }\textbf {\bibinfo {volume} {1761}},\ \bibinfo
  {pages} {897} (\bibinfo {year} {2006})}\BibitemShut {NoStop}%
\bibitem [{\citenamefont {Mim}\ and\ \citenamefont {Unger}(2012)}]{mim12a}%
  \BibitemOpen
  \bibfield  {author} {\bibinfo {author} {\bibfnamefont {C.}~\bibnamefont
  {Mim}}\ and\ \bibinfo {author} {\bibfnamefont {V.~M.}\ \bibnamefont
  {Unger}},\ }\href@noop {} {\bibfield  {journal} {\bibinfo  {journal} {Trends
  Biochem. Sci.}\ }\textbf {\bibinfo {volume} {37}},\ \bibinfo {pages} {526}
  (\bibinfo {year} {2012})}\BibitemShut {NoStop}%
\bibitem [{\citenamefont {Baumgart}\ \emph {et~al.}(2011)\citenamefont
  {Baumgart}, \citenamefont {Capraro}, \citenamefont {Zhu},\ and\ \citenamefont
  {Das}}]{baum11}%
  \BibitemOpen
  \bibfield  {author} {\bibinfo {author} {\bibfnamefont {T.}~\bibnamefont
  {Baumgart}}, \bibinfo {author} {\bibfnamefont {B.~R.}\ \bibnamefont
  {Capraro}}, \bibinfo {author} {\bibfnamefont {C.}~\bibnamefont {Zhu}}, \ and\
  \bibinfo {author} {\bibfnamefont {S.~L.}\ \bibnamefont {Das}},\ }\href@noop
  {} {\bibfield  {journal} {\bibinfo  {journal} {Annu. Rev. Phys. Chem.}\
  }\textbf {\bibinfo {volume} {62}},\ \bibinfo {pages} {483} (\bibinfo {year}
  {2011})}\BibitemShut {NoStop}%
\bibitem [{\citenamefont {Sorre}\ \emph {et~al.}(2012)\citenamefont {Sorre},
  \citenamefont {Callan-Jones}, \citenamefont {Manzi}, \citenamefont {Goud},
  \citenamefont {Prost}, \citenamefont {Bassereau},\ and\ \citenamefont
  {Roux}}]{sorr12}%
  \BibitemOpen
  \bibfield  {author} {\bibinfo {author} {\bibfnamefont {B.}~\bibnamefont
  {Sorre}}, \bibinfo {author} {\bibfnamefont {A.}~\bibnamefont {Callan-Jones}},
  \bibinfo {author} {\bibfnamefont {J.}~\bibnamefont {Manzi}}, \bibinfo
  {author} {\bibfnamefont {B.}~\bibnamefont {Goud}}, \bibinfo {author}
  {\bibfnamefont {J.}~\bibnamefont {Prost}}, \bibinfo {author} {\bibfnamefont
  {P.}~\bibnamefont {Bassereau}}, \ and\ \bibinfo {author} {\bibfnamefont
  {A.}~\bibnamefont {Roux}},\ }\href@noop {} {\bibfield  {journal} {\bibinfo
  {journal} {Proc.\ Natl.\ Acad.\ Sci.\ USA}\ }\textbf {\bibinfo {volume}
  {109}},\ \bibinfo {pages} {173} (\bibinfo {year} {2012})}\BibitemShut
  {NoStop}%
\bibitem [{\citenamefont {Pr{\'e}vost}\ \emph {et~al.}(2015)\citenamefont
  {Pr{\'e}vost}, \citenamefont {Zhao}, \citenamefont {Manzi}, \citenamefont
  {Lemichez}, \citenamefont {Lappalainen}, \citenamefont {Callan-Jones},\ and\
  \citenamefont {Bassereau}}]{prev15}%
  \BibitemOpen
  \bibfield  {author} {\bibinfo {author} {\bibfnamefont {C.}~\bibnamefont
  {Pr{\'e}vost}}, \bibinfo {author} {\bibfnamefont {H.}~\bibnamefont {Zhao}},
  \bibinfo {author} {\bibfnamefont {J.}~\bibnamefont {Manzi}}, \bibinfo
  {author} {\bibfnamefont {E.}~\bibnamefont {Lemichez}}, \bibinfo {author}
  {\bibfnamefont {P.}~\bibnamefont {Lappalainen}}, \bibinfo {author}
  {\bibfnamefont {A.}~\bibnamefont {Callan-Jones}}, \ and\ \bibinfo {author}
  {\bibfnamefont {P.}~\bibnamefont {Bassereau}},\ }\href@noop {} {\bibfield
  {journal} {\bibinfo  {journal} {Nat. Commun.}\ }\textbf {\bibinfo {volume}
  {6}},\ \bibinfo {pages} {8529} (\bibinfo {year} {2015})}\BibitemShut
  {NoStop}%
\bibitem [{\citenamefont {Rosholm}\ \emph {et~al.}(2017)\citenamefont
  {Rosholm}, \citenamefont {Leijnse}, \citenamefont {Mantsiou}, \citenamefont
  {Tkach}, \citenamefont {Pedersen}, \citenamefont {Wirth}, \citenamefont
  {Oddershede}, \citenamefont {Jensen}, \citenamefont {Martinez}, \citenamefont
  {Hatzakis}, \citenamefont {Bendix}, \citenamefont {Callan-Jones},\ and\
  \citenamefont {Stamou}}]{rosh17}%
  \BibitemOpen
  \bibfield  {author} {\bibinfo {author} {\bibfnamefont {K.~R.}\ \bibnamefont
  {Rosholm}}, \bibinfo {author} {\bibfnamefont {N.}~\bibnamefont {Leijnse}},
  \bibinfo {author} {\bibfnamefont {A.}~\bibnamefont {Mantsiou}}, \bibinfo
  {author} {\bibfnamefont {V.}~\bibnamefont {Tkach}}, \bibinfo {author}
  {\bibfnamefont {S.~L.}\ \bibnamefont {Pedersen}}, \bibinfo {author}
  {\bibfnamefont {V.~F.}\ \bibnamefont {Wirth}}, \bibinfo {author}
  {\bibfnamefont {L.~B.}\ \bibnamefont {Oddershede}}, \bibinfo {author}
  {\bibfnamefont {K.~J.}\ \bibnamefont {Jensen}}, \bibinfo {author}
  {\bibfnamefont {K.~L.}\ \bibnamefont {Martinez}}, \bibinfo {author}
  {\bibfnamefont {N.~S.}\ \bibnamefont {Hatzakis}}, \bibinfo {author}
  {\bibfnamefont {P.~M.}\ \bibnamefont {Bendix}}, \bibinfo {author}
  {\bibfnamefont {A.}~\bibnamefont {Callan-Jones}}, \ and\ \bibinfo {author}
  {\bibfnamefont {D.}~\bibnamefont {Stamou}},\ }\href@noop {} {\bibfield
  {journal} {\bibinfo  {journal} {Nat. Chem. Biol.}\ }\textbf {\bibinfo
  {volume} {13}},\ \bibinfo {pages} {724} (\bibinfo {year} {2017})}\BibitemShut
  {NoStop}%
\bibitem [{\citenamefont {Zeno}\ \emph {et~al.}(2019)\citenamefont {Zeno},
  \citenamefont {Snead}, \citenamefont {Thatte},\ and\ \citenamefont
  {Stachowiak}}]{zeno19}%
  \BibitemOpen
  \bibfield  {author} {\bibinfo {author} {\bibfnamefont {W.~F.}\ \bibnamefont
  {Zeno}}, \bibinfo {author} {\bibfnamefont {W.~T.}\ \bibnamefont {Snead}},
  \bibinfo {author} {\bibfnamefont {A.~S.}\ \bibnamefont {Thatte}}, \ and\
  \bibinfo {author} {\bibfnamefont {J.~C.}\ \bibnamefont {Stachowiak}},\
  }\href@noop {} {\bibfield  {journal} {\bibinfo  {journal} {Soft Matter}\
  }\textbf {\bibinfo {volume} {15}},\ \bibinfo {pages} {8706} (\bibinfo {year}
  {2019})}\BibitemShut {NoStop}%
\bibitem [{\citenamefont {{Sunil Kumar}}\ \emph {et~al.}(2001)\citenamefont
  {{Sunil Kumar}}, \citenamefont {Gompper},\ and\ \citenamefont
  {Lipowsky}}]{kuma01}%
  \BibitemOpen
  \bibfield  {author} {\bibinfo {author} {\bibfnamefont {P.~B.}\ \bibnamefont
  {{Sunil Kumar}}}, \bibinfo {author} {\bibfnamefont {G.}~\bibnamefont
  {Gompper}}, \ and\ \bibinfo {author} {\bibfnamefont {R.}~\bibnamefont
  {Lipowsky}},\ }\href@noop {} {\bibfield  {journal} {\bibinfo  {journal}
  {Phys.\ Rev.\ Lett.}\ }\textbf {\bibinfo {volume} {86}},\ \bibinfo {pages}
  {3911} (\bibinfo {year} {2001})}\BibitemShut {NoStop}%
\bibitem [{\citenamefont {Kohyama}\ \emph {et~al.}(2003)\citenamefont
  {Kohyama}, \citenamefont {Kroll},\ and\ \citenamefont {Gompper}}]{kohy03}%
  \BibitemOpen
  \bibfield  {author} {\bibinfo {author} {\bibfnamefont {T.}~\bibnamefont
  {Kohyama}}, \bibinfo {author} {\bibfnamefont {D.~M.}\ \bibnamefont {Kroll}},
  \ and\ \bibinfo {author} {\bibfnamefont {G.}~\bibnamefont {Gompper}},\
  }\href@noop {} {\bibfield  {journal} {\bibinfo  {journal} {Phys.\ Rev.\ E}\
  }\textbf {\bibinfo {volume} {68}},\ \bibinfo {pages} {061905} (\bibinfo
  {year} {2003})}\BibitemShut {NoStop}%
\bibitem [{\citenamefont {Noguchi}\ \emph {et~al.}(2015)\citenamefont
  {Noguchi}, \citenamefont {Sakashita},\ and\ \citenamefont {Imai}}]{nogu15a}%
  \BibitemOpen
  \bibfield  {author} {\bibinfo {author} {\bibfnamefont {H.}~\bibnamefont
  {Noguchi}}, \bibinfo {author} {\bibfnamefont {A.}~\bibnamefont {Sakashita}},
  \ and\ \bibinfo {author} {\bibfnamefont {M.}~\bibnamefont {Imai}},\
  }\href@noop {} {\bibfield  {journal} {\bibinfo  {journal} {Soft Matter}\
  }\textbf {\bibinfo {volume} {11}},\ \bibinfo {pages} {193} (\bibinfo {year}
  {2015})}\BibitemShut {NoStop}%
\bibitem [{\citenamefont {Pezeshkian}\ and\ \citenamefont
  {Ipsen}(2019)}]{peze19}%
  \BibitemOpen
  \bibfield  {author} {\bibinfo {author} {\bibfnamefont {W.}~\bibnamefont
  {Pezeshkian}}\ and\ \bibinfo {author} {\bibfnamefont {J.~H.}\ \bibnamefont
  {Ipsen}},\ }\href@noop {} {\bibfield  {journal} {\bibinfo  {journal} {Soft
  Matter}\ }\textbf {\bibinfo {volume} {15}},\ \bibinfo {pages} {9974}
  (\bibinfo {year} {2019})}\BibitemShut {NoStop}%
\bibitem [{\citenamefont {Tamemoto}\ and\ \citenamefont
  {Noguchi}(2020)}]{tame20}%
  \BibitemOpen
  \bibfield  {author} {\bibinfo {author} {\bibfnamefont {N.}~\bibnamefont
  {Tamemoto}}\ and\ \bibinfo {author} {\bibfnamefont {H.}~\bibnamefont
  {Noguchi}},\ }\href@noop {} {\bibfield  {journal} {\bibinfo  {journal} {Sci.
  Rep.}\ }\textbf {\bibinfo {volume} {10}},\ \bibinfo {pages} {19582} (\bibinfo
  {year} {2020})}\BibitemShut {NoStop}%
\bibitem [{\citenamefont {Noguchi}(2017)}]{nogu17a}%
  \BibitemOpen
  \bibfield  {author} {\bibinfo {author} {\bibfnamefont {H.}~\bibnamefont
  {Noguchi}},\ }\href@noop {} {\bibfield  {journal} {\bibinfo  {journal} {Soft\
  Matter}\ }\textbf {\bibinfo {volume} {13}},\ \bibinfo {pages} {7771}
  (\bibinfo {year} {2017})}\BibitemShut {NoStop}%
\bibitem [{\citenamefont {Bagatolli}\ and\ \citenamefont
  {Kumar}(2009)}]{baga09}%
  \BibitemOpen
  \bibfield  {author} {\bibinfo {author} {\bibfnamefont {L.}~\bibnamefont
  {Bagatolli}}\ and\ \bibinfo {author} {\bibfnamefont {P.~B.~S.}\ \bibnamefont
  {Kumar}},\ }\href@noop {} {\bibfield  {journal} {\bibinfo  {journal} {Soft\
  Matter}\ }\textbf {\bibinfo {volume} {5}},\ \bibinfo {pages} {3234} (\bibinfo
  {year} {2009})}\BibitemShut {NoStop}%
\bibitem [{\citenamefont {Nakagawa}\ and\ \citenamefont
  {Noguchi}(2018)}]{naka18}%
  \BibitemOpen
  \bibfield  {author} {\bibinfo {author} {\bibfnamefont {K.~M.}\ \bibnamefont
  {Nakagawa}}\ and\ \bibinfo {author} {\bibfnamefont {H.}~\bibnamefont
  {Noguchi}},\ }\href@noop {} {\bibfield  {journal} {\bibinfo  {journal} {Soft
  Matter}\ }\textbf {\bibinfo {volume} {14}},\ \bibinfo {pages} {1397}
  (\bibinfo {year} {2018})}\BibitemShut {NoStop}%
\bibitem [{\citenamefont {Lowengrub}\ \emph {et~al.}(2009)\citenamefont
  {Lowengrub}, \citenamefont {R\"atz},\ and\ \citenamefont {Voigt}}]{lowe09}%
  \BibitemOpen
  \bibfield  {author} {\bibinfo {author} {\bibfnamefont {J.~S.}\ \bibnamefont
  {Lowengrub}}, \bibinfo {author} {\bibfnamefont {A.}~\bibnamefont {R\"atz}}, \
  and\ \bibinfo {author} {\bibfnamefont {A.}~\bibnamefont {Voigt}},\
  }\href@noop {} {\bibfield  {journal} {\bibinfo  {journal} {Phys. Rev. E}\
  }\textbf {\bibinfo {volume} {79}},\ \bibinfo {pages} {031926} (\bibinfo
  {year} {2009})}\BibitemShut {NoStop}%
\bibitem [{\citenamefont {Noguchi}(2016)}]{nogu16}%
  \BibitemOpen
  \bibfield  {author} {\bibinfo {author} {\bibfnamefont {H.}~\bibnamefont
  {Noguchi}},\ }\href@noop {} {\bibfield  {journal} {\bibinfo  {journal} {Sci.\
  Rep.}\ }\textbf {\bibinfo {volume} {6}},\ \bibinfo {pages} {20935} (\bibinfo
  {year} {2016})}\BibitemShut {NoStop}%
\bibitem [{\citenamefont {Noguchi}(2019)}]{nogu19a}%
  \BibitemOpen
  \bibfield  {author} {\bibinfo {author} {\bibfnamefont {H.}~\bibnamefont
  {Noguchi}},\ }\href@noop {} {\bibfield  {journal} {\bibinfo  {journal} {Sci.\
  Rep.}\ }\textbf {\bibinfo {volume} {9}},\ \bibinfo {pages} {11721} (\bibinfo
  {year} {2019})}\BibitemShut {NoStop}%
\bibitem [{\citenamefont {Ramakrishnan}\ \emph {et~al.}(2018)\citenamefont
  {Ramakrishnan}, \citenamefont {Bradley}, \citenamefont {Tourdot},\ and\
  \citenamefont {Radhakrishnan}}]{rama18}%
  \BibitemOpen
  \bibfield  {author} {\bibinfo {author} {\bibfnamefont {N.}~\bibnamefont
  {Ramakrishnan}}, \bibinfo {author} {\bibfnamefont {R.~P.}\ \bibnamefont
  {Bradley}}, \bibinfo {author} {\bibfnamefont {R.~W.}\ \bibnamefont
  {Tourdot}}, \ and\ \bibinfo {author} {\bibfnamefont {R.}~\bibnamefont
  {Radhakrishnan}},\ }\href@noop {} {\bibfield  {journal} {\bibinfo  {journal}
  {J. Phys.: Condens. Matter}\ }\textbf {\bibinfo {volume} {30}},\ \bibinfo
  {pages} {273001} (\bibinfo {year} {2018})}\BibitemShut {NoStop}%
\bibitem [{\citenamefont {Lipowsky}(1992)}]{lipo92}%
  \BibitemOpen
  \bibfield  {author} {\bibinfo {author} {\bibfnamefont {R.}~\bibnamefont
  {Lipowsky}},\ }\href@noop {} {\bibfield  {journal} {\bibinfo  {journal} {J.
  Phys. II France}\ }\textbf {\bibinfo {volume} {2}},\ \bibinfo {pages} {1825}
  (\bibinfo {year} {1992})}\BibitemShut {NoStop}%
\bibitem [{\citenamefont {Sens}(2004)}]{sens03}%
  \BibitemOpen
  \bibfield  {author} {\bibinfo {author} {\bibfnamefont {P.}~\bibnamefont
  {Sens}},\ }\href@noop {} {\bibfield  {journal} {\bibinfo  {journal} {Phys.
  Rev. Lett.}\ }\textbf {\bibinfo {volume} {93}},\ \bibinfo {pages} {108103}
  (\bibinfo {year} {2004})}\BibitemShut {NoStop}%
\bibitem [{\citenamefont {Foret}(2014)}]{fore14}%
  \BibitemOpen
  \bibfield  {author} {\bibinfo {author} {\bibfnamefont {L.}~\bibnamefont
  {Foret}},\ }\href@noop {} {\bibfield  {journal} {\bibinfo  {journal} {Eur.
  Phys. J. E}\ }\textbf {\bibinfo {volume} {37}},\ \bibinfo {pages} {42}
  (\bibinfo {year} {2014})}\BibitemShut {NoStop}%
\bibitem [{\citenamefont {Frey}\ and\ \citenamefont {Schwarz}(2020)}]{frey20}%
  \BibitemOpen
  \bibfield  {author} {\bibinfo {author} {\bibfnamefont {F.}~\bibnamefont
  {Frey}}\ and\ \bibinfo {author} {\bibfnamefont {U.~S.}\ \bibnamefont
  {Schwarz}},\ }\href@noop {} {\bibfield  {journal} {\bibinfo  {journal} {Soft
  Matter}\ }\textbf {\bibinfo {volume} {16}},\ \bibinfo {pages} {10723}
  (\bibinfo {year} {2020})}\BibitemShut {NoStop}%
\bibitem [{\citenamefont {Canham}(1970)}]{canh70}%
  \BibitemOpen
  \bibfield  {author} {\bibinfo {author} {\bibfnamefont {P.~B.}\ \bibnamefont
  {Canham}},\ }\href@noop {} {\bibfield  {journal} {\bibinfo  {journal} {J.
  Theor. Biol.}\ }\textbf {\bibinfo {volume} {26}},\ \bibinfo {pages} {61}
  (\bibinfo {year} {1970})}\BibitemShut {NoStop}%
\bibitem [{\citenamefont {Helfrich}(1973)}]{helf73}%
  \BibitemOpen
  \bibfield  {author} {\bibinfo {author} {\bibfnamefont {W.}~\bibnamefont
  {Helfrich}},\ }\href@noop {} {\bibfield  {journal} {\bibinfo  {journal} {Z.\
  Naturforsch}\ }\textbf {\bibinfo {volume} {28c}},\ \bibinfo {pages} {693}
  (\bibinfo {year} {1973})}\BibitemShut {NoStop}%
\bibitem [{\citenamefont {Safran}(1994)}]{safr94}%
  \BibitemOpen
  \bibfield  {author} {\bibinfo {author} {\bibfnamefont {S.~A.}\ \bibnamefont
  {Safran}},\ }\href@noop {} {\emph {\bibinfo {title} {Statistical
  Thermodynamics of Surfaces, Interfaces, and Membranes}}}\ (\bibinfo
  {publisher} {Addison-Wesley},\ \bibinfo {address} {Reading, MA},\ \bibinfo
  {year} {1994})\BibitemShut {NoStop}%
\bibitem [{\citenamefont {Goutaland}\ \emph {et~al.}(2021)\citenamefont
  {Goutaland}, \citenamefont {van Wijland}, \citenamefont {Fournier},\ and\
  \citenamefont {Noguchi}}]{gout21}%
  \BibitemOpen
  \bibfield  {author} {\bibinfo {author} {\bibfnamefont {Q.}~\bibnamefont
  {Goutaland}}, \bibinfo {author} {\bibfnamefont {F.}~\bibnamefont {van
  Wijland}}, \bibinfo {author} {\bibfnamefont {J.-B.}\ \bibnamefont
  {Fournier}}, \ and\ \bibinfo {author} {\bibfnamefont {H.}~\bibnamefont
  {Noguchi}},\ }\href@noop {} {\bibfield  {journal} {\bibinfo  {journal} {Soft
  Matter}\ }\textbf {\bibinfo {volume} {17}},\ \bibinfo {pages} {5560}
  (\bibinfo {year} {2021})}\BibitemShut {NoStop}%
\bibitem [{\citenamefont {Noguchi}(2015)}]{nogu15b}%
  \BibitemOpen
  \bibfield  {author} {\bibinfo {author} {\bibfnamefont {H.}~\bibnamefont
  {Noguchi}},\ }\href@noop {} {\bibfield  {journal} {\bibinfo  {journal} {J.
  Chem. Phys.}\ }\textbf {\bibinfo {volume} {143}},\ \bibinfo {pages} {243109}
  (\bibinfo {year} {2015})}\BibitemShut {NoStop}%
\bibitem [{\citenamefont {Reynwar}\ \emph {et~al.}(2007)\citenamefont
  {Reynwar}, \citenamefont {Ilya}, \citenamefont {Harmandaris}, \citenamefont
  {M{\"u}ller}, \citenamefont {Kremer},\ and\ \citenamefont
  {Deserno}}]{reyn07}%
  \BibitemOpen
  \bibfield  {author} {\bibinfo {author} {\bibfnamefont {B.~J.}\ \bibnamefont
  {Reynwar}}, \bibinfo {author} {\bibfnamefont {G.}~\bibnamefont {Ilya}},
  \bibinfo {author} {\bibfnamefont {V.~A.}\ \bibnamefont {Harmandaris}},
  \bibinfo {author} {\bibfnamefont {M.~M.}\ \bibnamefont {M{\"u}ller}},
  \bibinfo {author} {\bibfnamefont {K.}~\bibnamefont {Kremer}}, \ and\ \bibinfo
  {author} {\bibfnamefont {M.}~\bibnamefont {Deserno}},\ }\href@noop {}
  {\bibfield  {journal} {\bibinfo  {journal} {Nature}\ }\textbf {\bibinfo
  {volume} {447}},\ \bibinfo {pages} {461} (\bibinfo {year}
  {2007})}\BibitemShut {NoStop}%
\bibitem [{\citenamefont {Auth}\ and\ \citenamefont {Gompper}(2009)}]{auth09}%
  \BibitemOpen
  \bibfield  {author} {\bibinfo {author} {\bibfnamefont {T.}~\bibnamefont
  {Auth}}\ and\ \bibinfo {author} {\bibfnamefont {G.}~\bibnamefont {Gompper}},\
  }\href@noop {} {\bibfield  {journal} {\bibinfo  {journal} {Phys. Rev. E}\
  }\textbf {\bibinfo {volume} {80}},\ \bibinfo {pages} {031901} (\bibinfo
  {year} {2009})}\BibitemShut {NoStop}%
\bibitem [{\citenamefont {Aranda-Espinoza}\ \emph {et~al.}(1996)\citenamefont
  {Aranda-Espinoza}, \citenamefont {Berman}, \citenamefont {Dan}, \citenamefont
  {Pincus},\ and\ \citenamefont {Safran}}]{aran96}%
  \BibitemOpen
  \bibfield  {author} {\bibinfo {author} {\bibfnamefont {H.}~\bibnamefont
  {Aranda-Espinoza}}, \bibinfo {author} {\bibfnamefont {A.}~\bibnamefont
  {Berman}}, \bibinfo {author} {\bibfnamefont {N.}~\bibnamefont {Dan}},
  \bibinfo {author} {\bibfnamefont {P.}~\bibnamefont {Pincus}}, \ and\ \bibinfo
  {author} {\bibfnamefont {S.}~\bibnamefont {Safran}},\ }\href@noop {}
  {\bibfield  {journal} {\bibinfo  {journal} {Biophys. J.}\ }\textbf {\bibinfo
  {volume} {71}},\ \bibinfo {pages} {648} (\bibinfo {year} {1996})}\BibitemShut
  {NoStop}%
\bibitem [{\citenamefont {Fournier}(1999)}]{four99}%
  \BibitemOpen
  \bibfield  {author} {\bibinfo {author} {\bibfnamefont {J.-B.}\ \bibnamefont
  {Fournier}},\ }\href@noop {} {\bibfield  {journal} {\bibinfo  {journal}
  {Eur.\ Phys.\ J. B}\ }\textbf {\bibinfo {volume} {11}},\ \bibinfo {pages}
  {261} (\bibinfo {year} {1999})}\BibitemShut {NoStop}%
\bibitem [{\citenamefont {Noguchi}\ and\ \citenamefont
  {Fournier}(2017)}]{nogu17}%
  \BibitemOpen
  \bibfield  {author} {\bibinfo {author} {\bibfnamefont {H.}~\bibnamefont
  {Noguchi}}\ and\ \bibinfo {author} {\bibfnamefont {J.-B.}\ \bibnamefont
  {Fournier}},\ }\href@noop {} {\bibfield  {journal} {\bibinfo  {journal}
  {Soft\ Matter}\ }\textbf {\bibinfo {volume} {13}},\ \bibinfo {pages} {4099}
  (\bibinfo {year} {2017})}\BibitemShut {NoStop}%
\bibitem [{\citenamefont {Sachin~Krishnan}\ \emph {et~al.}(2019)\citenamefont
  {Sachin~Krishnan}, \citenamefont {Das},\ and\ \citenamefont
  {Sunil~Kumar}}]{kris19}%
  \BibitemOpen
  \bibfield  {author} {\bibinfo {author} {\bibfnamefont {T.~V.}\ \bibnamefont
  {Sachin~Krishnan}}, \bibinfo {author} {\bibfnamefont {S.~L.}\ \bibnamefont
  {Das}}, \ and\ \bibinfo {author} {\bibfnamefont {P.~B.}\ \bibnamefont
  {Sunil~Kumar}},\ }\href@noop {} {\bibfield  {journal} {\bibinfo  {journal}
  {Soft Matter}\ }\textbf {\bibinfo {volume} {15}},\ \bibinfo {pages} {2071}
  (\bibinfo {year} {2019})}\BibitemShut {NoStop}%
\bibitem [{\citenamefont {Shi}\ and\ \citenamefont {Baumgart}(2015)}]{shi15}%
  \BibitemOpen
  \bibfield  {author} {\bibinfo {author} {\bibfnamefont {Z.}~\bibnamefont
  {Shi}}\ and\ \bibinfo {author} {\bibfnamefont {T.}~\bibnamefont {Baumgart}},\
  }\href@noop {} {\bibfield  {journal} {\bibinfo  {journal} {Nat. Commun.}\
  }\textbf {\bibinfo {volume} {6}},\ \bibinfo {pages} {5974} (\bibinfo {year}
  {2015})}\BibitemShut {NoStop}%
\bibitem [{\citenamefont {Gov}(2018)}]{gov18}%
  \BibitemOpen
  \bibfield  {author} {\bibinfo {author} {\bibfnamefont {N.~S.}\ \bibnamefont
  {Gov}},\ }\href@noop {} {\bibfield  {journal} {\bibinfo  {journal} {Phil.
  Trans. R. Soc. B}\ }\textbf {\bibinfo {volume} {373}},\ \bibinfo {pages}
  {20170115} (\bibinfo {year} {2018})}\BibitemShut {NoStop}%
\bibitem [{\citenamefont {Tozzi}\ \emph {et~al.}(2019)\citenamefont {Tozzi},
  \citenamefont {Walani},\ and\ \citenamefont {Arroyo}}]{tozz19}%
  \BibitemOpen
  \bibfield  {author} {\bibinfo {author} {\bibfnamefont {C.}~\bibnamefont
  {Tozzi}}, \bibinfo {author} {\bibfnamefont {N.}~\bibnamefont {Walani}}, \
  and\ \bibinfo {author} {\bibfnamefont {M.}~\bibnamefont {Arroyo}},\
  }\href@noop {} {\bibfield  {journal} {\bibinfo  {journal} {New J. Phys.}\
  }\textbf {\bibinfo {volume} {21}},\ \bibinfo {pages} {093004} (\bibinfo
  {year} {2019})}\BibitemShut {NoStop}%
\bibitem [{\citenamefont {Ramaswamy}\ \emph {et~al.}(2000)\citenamefont
  {Ramaswamy}, \citenamefont {Toner},\ and\ \citenamefont {Prost}}]{rama00}%
  \BibitemOpen
  \bibfield  {author} {\bibinfo {author} {\bibfnamefont {S.}~\bibnamefont
  {Ramaswamy}}, \bibinfo {author} {\bibfnamefont {J.}~\bibnamefont {Toner}}, \
  and\ \bibinfo {author} {\bibfnamefont {J.}~\bibnamefont {Prost}},\
  }\href@noop {} {\bibfield  {journal} {\bibinfo  {journal} {Phys. Rev. Lett.}\
  }\textbf {\bibinfo {volume} {84}},\ \bibinfo {pages} {3494} (\bibinfo {year}
  {2000})}\BibitemShut {NoStop}%
\bibitem [{\citenamefont {Shlomovitz}\ and\ \citenamefont
  {Gov}(2009)}]{shlo09}%
  \BibitemOpen
  \bibfield  {author} {\bibinfo {author} {\bibfnamefont {R.}~\bibnamefont
  {Shlomovitz}}\ and\ \bibinfo {author} {\bibfnamefont {N.~S.}\ \bibnamefont
  {Gov}},\ }\href@noop {} {\bibfield  {journal} {\bibinfo  {journal} {Phys.
  Biol.}\ }\textbf {\bibinfo {volume} {6}},\ \bibinfo {pages} {046017}
  (\bibinfo {year} {2009})}\BibitemShut {NoStop}%
\bibitem [{\citenamefont {Hassinger}\ \emph {et~al.}(2017)\citenamefont
  {Hassinger}, \citenamefont {Oster}, \citenamefont {Drubin},\ and\
  \citenamefont {Rangamani}}]{hass17}%
  \BibitemOpen
  \bibfield  {author} {\bibinfo {author} {\bibfnamefont {J.~E.}\ \bibnamefont
  {Hassinger}}, \bibinfo {author} {\bibfnamefont {G.}~\bibnamefont {Oster}},
  \bibinfo {author} {\bibfnamefont {D.~G.}\ \bibnamefont {Drubin}}, \ and\
  \bibinfo {author} {\bibfnamefont {P.}~\bibnamefont {Rangamani}},\ }\href@noop
  {} {\bibfield  {journal} {\bibinfo  {journal} {Proc.\ Natl.\ Acad.\ Sci.\
  USA}\ }\textbf {\bibinfo {volume} {114}},\ \bibinfo {pages} {E1118} (\bibinfo
  {year} {2017})}\BibitemShut {NoStop}%
\bibitem [{\citenamefont {Evans}\ and\ \citenamefont {Ludwig}(2000)}]{evan00}%
  \BibitemOpen
  \bibfield  {author} {\bibinfo {author} {\bibfnamefont {E.~A.}\ \bibnamefont
  {Evans}}\ and\ \bibinfo {author} {\bibfnamefont {F.}~\bibnamefont {Ludwig}},\
  }\href@noop {} {\bibfield  {journal} {\bibinfo  {journal} {J.\ Phys.\
  Condens.\ Matter}\ }\textbf {\bibinfo {volume} {12}},\ \bibinfo {pages}
  {A315} (\bibinfo {year} {2000})}\BibitemShut {NoStop}%
\bibitem [{\citenamefont {Evans}\ \emph {et~al.}(2003)\citenamefont {Evans},
  \citenamefont {Heinrich}, \citenamefont {Ludwig},\ and\ \citenamefont
  {Rawicz}}]{evan03}%
  \BibitemOpen
  \bibfield  {author} {\bibinfo {author} {\bibfnamefont {E.~A.}\ \bibnamefont
  {Evans}}, \bibinfo {author} {\bibfnamefont {V.}~\bibnamefont {Heinrich}},
  \bibinfo {author} {\bibfnamefont {F.}~\bibnamefont {Ludwig}}, \ and\ \bibinfo
  {author} {\bibfnamefont {W.}~\bibnamefont {Rawicz}},\ }\href@noop {}
  {\bibfield  {journal} {\bibinfo  {journal} {Biophys.\ J.}\ }\textbf {\bibinfo
  {volume} {85}},\ \bibinfo {pages} {2342} (\bibinfo {year}
  {2003})}\BibitemShut {NoStop}%
\bibitem [{\citenamefont {Ly}\ and\ \citenamefont {Longo}(2004)}]{ly04}%
  \BibitemOpen
  \bibfield  {author} {\bibinfo {author} {\bibfnamefont {H.~V.}\ \bibnamefont
  {Ly}}\ and\ \bibinfo {author} {\bibfnamefont {M.~L.}\ \bibnamefont {Longo}},\
  }\href@noop {} {\bibfield  {journal} {\bibinfo  {journal} {Biophys.\ J.}\
  }\textbf {\bibinfo {volume} {87}},\ \bibinfo {pages} {1013} (\bibinfo {year}
  {2004})}\BibitemShut {NoStop}%
\bibitem [{\citenamefont {Kornberg}\ and\ \citenamefont
  {McConnell}(1971)}]{korn71}%
  \BibitemOpen
  \bibfield  {author} {\bibinfo {author} {\bibfnamefont {R.~D.}\ \bibnamefont
  {Kornberg}}\ and\ \bibinfo {author} {\bibfnamefont {H.~M.}\ \bibnamefont
  {McConnell}},\ }\href@noop {} {\bibfield  {journal} {\bibinfo  {journal}
  {Biochemistry}\ }\textbf {\bibinfo {volume} {10}},\ \bibinfo {pages} {1111}
  (\bibinfo {year} {1971})}\BibitemShut {NoStop}%
\bibitem [{\citenamefont {Contreras}\ \emph {et~al.}(2009)\citenamefont
  {Contreras}, \citenamefont {S{\'a}nchez-Magraner}, \citenamefont {Alonso},\
  and\ \citenamefont {Go{\~n}i}}]{cont09}%
  \BibitemOpen
  \bibfield  {author} {\bibinfo {author} {\bibfnamefont {F.-X.}\ \bibnamefont
  {Contreras}}, \bibinfo {author} {\bibfnamefont {L.}~\bibnamefont
  {S{\'a}nchez-Magraner}}, \bibinfo {author} {\bibfnamefont {A.}~\bibnamefont
  {Alonso}}, \ and\ \bibinfo {author} {\bibfnamefont {F.~M.}\ \bibnamefont
  {Go{\~n}i}},\ }\href@noop {} {\bibfield  {journal} {\bibinfo  {journal} {FEBS
  Lett.}\ }\textbf {\bibinfo {volume} {584}},\ \bibinfo {pages} {1779}
  (\bibinfo {year} {2009})}\BibitemShut {NoStop}%
\bibitem [{\citenamefont {Hamilton}(2003)}]{hami03}%
  \BibitemOpen
  \bibfield  {author} {\bibinfo {author} {\bibfnamefont {J.~A.}\ \bibnamefont
  {Hamilton}},\ }\href@noop {} {\bibfield  {journal} {\bibinfo  {journal}
  {Curr. Opin. Lipidol.}\ }\textbf {\bibinfo {volume} {14}},\ \bibinfo {pages}
  {263} (\bibinfo {year} {2003})}\BibitemShut {NoStop}%
\bibitem [{\citenamefont {Steck}\ \emph {et~al.}(2002)\citenamefont {Steck},
  \citenamefont {Ye},\ and\ \citenamefont {Lange}}]{stec02}%
  \BibitemOpen
  \bibfield  {author} {\bibinfo {author} {\bibfnamefont {T.~L.}\ \bibnamefont
  {Steck}}, \bibinfo {author} {\bibfnamefont {J.}~\bibnamefont {Ye}}, \ and\
  \bibinfo {author} {\bibfnamefont {Y.}~\bibnamefont {Lange}},\ }\href@noop {}
  {\bibfield  {journal} {\bibinfo  {journal} {Biophys. J.}\ }\textbf {\bibinfo
  {volume} {83}},\ \bibinfo {pages} {2118} (\bibinfo {year}
  {2002})}\BibitemShut {NoStop}%
\bibitem [{\citenamefont {Bruckner}\ \emph {et~al.}(2009)\citenamefont
  {Bruckner}, \citenamefont {Mansy}, \citenamefont {Ricardo}, \citenamefont
  {Mahadevan},\ and\ \citenamefont {Szostak}}]{bruc09}%
  \BibitemOpen
  \bibfield  {author} {\bibinfo {author} {\bibfnamefont {R.~J.}\ \bibnamefont
  {Bruckner}}, \bibinfo {author} {\bibfnamefont {S.~S.}\ \bibnamefont {Mansy}},
  \bibinfo {author} {\bibfnamefont {A.}~\bibnamefont {Ricardo}}, \bibinfo
  {author} {\bibfnamefont {L.}~\bibnamefont {Mahadevan}}, \ and\ \bibinfo
  {author} {\bibfnamefont {J.~W.}\ \bibnamefont {Szostak}},\ }\href@noop {}
  {\bibfield  {journal} {\bibinfo  {journal} {Biophys. J.}\ }\textbf {\bibinfo
  {volume} {97}},\ \bibinfo {pages} {3113} (\bibinfo {year}
  {2009})}\BibitemShut {NoStop}%
\bibitem [{\citenamefont {Miettinen}\ and\ \citenamefont
  {Lipowsky}(2019)}]{miet19}%
  \BibitemOpen
  \bibfield  {author} {\bibinfo {author} {\bibfnamefont {M.~S.}\ \bibnamefont
  {Miettinen}}\ and\ \bibinfo {author} {\bibfnamefont {R.}~\bibnamefont
  {Lipowsky}},\ }\href@noop {} {\bibfield  {journal} {\bibinfo  {journal} {Nano
  Lett.}\ }\textbf {\bibinfo {volume} {19}},\ \bibinfo {pages} {5011} (\bibinfo
  {year} {2019})}\BibitemShut {NoStop}%
\bibitem [{\citenamefont {Kawaguchi}\ \emph {et~al.}(2020)\citenamefont
  {Kawaguchi}, \citenamefont {Nakagawa}, \citenamefont {Nakagawa},
  \citenamefont {Shindou}, \citenamefont {Nagao},\ and\ \citenamefont
  {Noguchi}}]{kawa20}%
  \BibitemOpen
  \bibfield  {author} {\bibinfo {author} {\bibfnamefont {K.}~\bibnamefont
  {Kawaguchi}}, \bibinfo {author} {\bibfnamefont {K.~M.}\ \bibnamefont
  {Nakagawa}}, \bibinfo {author} {\bibfnamefont {S.}~\bibnamefont {Nakagawa}},
  \bibinfo {author} {\bibfnamefont {H.}~\bibnamefont {Shindou}}, \bibinfo
  {author} {\bibfnamefont {H.}~\bibnamefont {Nagao}}, \ and\ \bibinfo {author}
  {\bibfnamefont {H.}~\bibnamefont {Noguchi}},\ }\href@noop {} {\bibfield
  {journal} {\bibinfo  {journal} {J. Chem. Phys.}\ }\textbf {\bibinfo {volume}
  {153}},\ \bibinfo {pages} {165101} (\bibinfo {year} {2020})}\BibitemShut
  {NoStop}%
\bibitem [{\citenamefont {Svetina}\ and\ \citenamefont
  {\v{Z}ek\v{s}}(1989)}]{svet89}%
  \BibitemOpen
  \bibfield  {author} {\bibinfo {author} {\bibfnamefont {S.}~\bibnamefont
  {Svetina}}\ and\ \bibinfo {author} {\bibfnamefont {B.}~\bibnamefont
  {\v{Z}ek\v{s}}},\ }\href {\doibase 10.1007/BF00257107} {\bibfield  {journal}
  {\bibinfo  {journal} {Euro. Biophys. J.}\ }\textbf {\bibinfo {volume} {17}},\
  \bibinfo {pages} {101} (\bibinfo {year} {1989})}\BibitemShut {NoStop}%
\bibitem [{\citenamefont {Busch}\ \emph {et~al.}(2015)\citenamefont {Busch},
  \citenamefont {Houser}, \citenamefont {Hayden}, \citenamefont {Sherman},
  \citenamefont {Lafer},\ and\ \citenamefont {Stachowiak}}]{busc15}%
  \BibitemOpen
  \bibfield  {author} {\bibinfo {author} {\bibfnamefont {D.~J.}\ \bibnamefont
  {Busch}}, \bibinfo {author} {\bibfnamefont {J.~R.}\ \bibnamefont {Houser}},
  \bibinfo {author} {\bibfnamefont {C.~C.}\ \bibnamefont {Hayden}}, \bibinfo
  {author} {\bibfnamefont {M.~B.}\ \bibnamefont {Sherman}}, \bibinfo {author}
  {\bibfnamefont {E.~M.}\ \bibnamefont {Lafer}}, \ and\ \bibinfo {author}
  {\bibfnamefont {J.~C.}\ \bibnamefont {Stachowiak}},\ }\href@noop {}
  {\bibfield  {journal} {\bibinfo  {journal} {Nat. Commun.}\ }\textbf {\bibinfo
  {volume} {6}},\ \bibinfo {pages} {7875} (\bibinfo {year} {2015})}\BibitemShut
  {NoStop}%
\bibitem [{\citenamefont {Hiergeist}\ and\ \citenamefont
  {Lipowsky}(1996)}]{hier96}%
  \BibitemOpen
  \bibfield  {author} {\bibinfo {author} {\bibfnamefont {C.}~\bibnamefont
  {Hiergeist}}\ and\ \bibinfo {author} {\bibfnamefont {R.}~\bibnamefont
  {Lipowsky}},\ }\href@noop {} {\bibfield  {journal} {\bibinfo  {journal} {J.
  Phys. II France}\ }\textbf {\bibinfo {volume} {6}},\ \bibinfo {pages} {1465}
  (\bibinfo {year} {1996})}\BibitemShut {NoStop}%
\bibitem [{\citenamefont {Bickel}\ and\ \citenamefont
  {Marques}(2006)}]{bick06}%
  \BibitemOpen
  \bibfield  {author} {\bibinfo {author} {\bibfnamefont {T.}~\bibnamefont
  {Bickel}}\ and\ \bibinfo {author} {\bibfnamefont {C.~M.}\ \bibnamefont
  {Marques}},\ }\href@noop {} {\bibfield  {journal} {\bibinfo  {journal} {J.
  Nanosci. Nanotechnol.}\ }\textbf {\bibinfo {volume} {6}},\ \bibinfo {pages}
  {2386} (\bibinfo {year} {2006})}\BibitemShut {NoStop}%
\bibitem [{\citenamefont {Wu}\ \emph {et~al.}(2013)\citenamefont {Wu},
  \citenamefont {Shiba},\ and\ \citenamefont {Noguchi}}]{wu13}%
  \BibitemOpen
  \bibfield  {author} {\bibinfo {author} {\bibfnamefont {H.}~\bibnamefont
  {Wu}}, \bibinfo {author} {\bibfnamefont {H.}~\bibnamefont {Shiba}}, \ and\
  \bibinfo {author} {\bibfnamefont {H.}~\bibnamefont {Noguchi}},\ }\href@noop
  {} {\bibfield  {journal} {\bibinfo  {journal} {Soft matter}\ }\textbf
  {\bibinfo {volume} {9}},\ \bibinfo {pages} {9907} (\bibinfo {year}
  {2013})}\BibitemShut {NoStop}%
\bibitem [{\citenamefont {Aimon}\ \emph {et~al.}(2014)\citenamefont {Aimon},
  \citenamefont {Callan-Jones}, \citenamefont {Berthaud}, \citenamefont
  {Pinot}, \citenamefont {Toombes},\ and\ \citenamefont {Bassereau}}]{aimo14}%
  \BibitemOpen
  \bibfield  {author} {\bibinfo {author} {\bibfnamefont {S.}~\bibnamefont
  {Aimon}}, \bibinfo {author} {\bibfnamefont {A.}~\bibnamefont {Callan-Jones}},
  \bibinfo {author} {\bibfnamefont {A.}~\bibnamefont {Berthaud}}, \bibinfo
  {author} {\bibfnamefont {M.}~\bibnamefont {Pinot}}, \bibinfo {author}
  {\bibfnamefont {G.~E.}\ \bibnamefont {Toombes}}, \ and\ \bibinfo {author}
  {\bibfnamefont {P.}~\bibnamefont {Bassereau}},\ }\href@noop {} {\bibfield
  {journal} {\bibinfo  {journal} {Dev. Cell}\ }\textbf {\bibinfo {volume}
  {28}},\ \bibinfo {pages} {212} (\bibinfo {year} {2014})}\BibitemShut
  {NoStop}%
\bibitem [{\citenamefont {Andreeva}\ \emph {et~al.}(2007)\citenamefont
  {Andreeva}, \citenamefont {Nesterenko}, \citenamefont {Fomkina},
  \citenamefont {Ternovsky}, \citenamefont {Suzina}, \citenamefont {Bakulina},
  \citenamefont {Solonin},\ and\ \citenamefont {Sineva}}]{andr07}%
  \BibitemOpen
  \bibfield  {author} {\bibinfo {author} {\bibfnamefont {Z.~I.}\ \bibnamefont
  {Andreeva}}, \bibinfo {author} {\bibfnamefont {V.~F.}\ \bibnamefont
  {Nesterenko}}, \bibinfo {author} {\bibfnamefont {M.~G.}\ \bibnamefont
  {Fomkina}}, \bibinfo {author} {\bibfnamefont {V.~I.}\ \bibnamefont
  {Ternovsky}}, \bibinfo {author} {\bibfnamefont {N.~E.}\ \bibnamefont
  {Suzina}}, \bibinfo {author} {\bibfnamefont {A.~Y.}\ \bibnamefont
  {Bakulina}}, \bibinfo {author} {\bibfnamefont {A.~S.}\ \bibnamefont
  {Solonin}}, \ and\ \bibinfo {author} {\bibfnamefont {E.~V.}\ \bibnamefont
  {Sineva}},\ }\href@noop {} {\bibfield  {journal} {\bibinfo  {journal}
  {Biochim.\ Biophys.\ Acta}\ }\textbf {\bibinfo {volume} {1768}},\ \bibinfo
  {pages} {253} (\bibinfo {year} {2007})}\BibitemShut {NoStop}%
\bibitem [{\citenamefont {Tozzi}\ \emph {et~al.}(2021)\citenamefont {Tozzi},
  \citenamefont {Walani}, \citenamefont {Roux}, \citenamefont {Roca-Cusachs},\
  and\ \citenamefont {Arroyo}}]{tozz21}%
  \BibitemOpen
  \bibfield  {author} {\bibinfo {author} {\bibfnamefont {C.}~\bibnamefont
  {Tozzi}}, \bibinfo {author} {\bibfnamefont {N.}~\bibnamefont {Walani}},
  \bibinfo {author} {\bibfnamefont {A.-L.~L.}\ \bibnamefont {Roux}}, \bibinfo
  {author} {\bibfnamefont {P.}~\bibnamefont {Roca-Cusachs}}, \ and\ \bibinfo
  {author} {\bibfnamefont {M.}~\bibnamefont {Arroyo}},\ }\href@noop {}
  {\bibfield  {journal} {\bibinfo  {journal} {Soft Matter}\ }\textbf {\bibinfo
  {volume} {17}},\ \bibinfo {pages} {3367} (\bibinfo {year}
  {2021})}\BibitemShut {NoStop}%
\end{thebibliography}
\end{document}